\documentclass[prd,twocolumn,twoside,preprintnumbers,superscriptaddress,nofootinbib]{revtex4}

\usepackage{amsmath}
\usepackage{amssymb}
\usepackage{amsfonts}
\usepackage{epsfig}
\usepackage{graphics}
\usepackage{graphicx}
\usepackage{hyperref}
\usepackage{multirow}
\usepackage{slashed}
\usepackage{wasysym}
\usepackage{xcolor}

\newcommand{\Eq}[1]{Eq.~\eqref{#1}}

\newcommand{\GeV}{\,\text{GeV}}

\newcommand{\be}{\begin{equation}}
\newcommand{\ee}{\end{equation}}
\newcommand{\vev}[1]{\left\langle #1 \right\rangle}
\newcommand{\ZZ}{\mathbb{Z}}
\newcommand{\Ref}[1]{Ref.~\cite{#1}}
\newcommand{\Eqs}[2]{Eqs.~\eqref{#1} and \eqref{#2}}
\newcommand{\bs}[1]{\boldsymbol #1}

\begin{document}

\preprint{\vbox{\hbox{SCIPP 16/04}}}

\title{Dark Matter Inelastic Up-Scattering with the Interstellar Plasma:\\ An Exciting New Source of X-Ray Lines, including at 3.5 keV}

\author{Francesco D'Eramo}
\email{fderamo@ucsc.edu}
\affiliation{Department of Physics, University of California, 1156 High St., Santa Cruz, CA 95064, USA}
\affiliation{Santa Cruz Institute for Particle Physics, Santa Cruz, CA 95064, USA} 
\author{Kevin Hambleton}
\email{khamblet@ucsc.edu}\affiliation{Department of Physics, University of California, 1156 High St., Santa Cruz, CA 95064, USA}
\author{Stefano Profumo}
\email{profumo@ucsc.edu}\affiliation{Department of Physics, University of California, 1156 High St., Santa Cruz, CA 95064, USA}\affiliation{Santa Cruz Institute for Particle Physics, Santa Cruz, CA 95064, USA} 
\author{Tim Stefaniak}
\email{tistefan@ucsc.edu}\affiliation{Department of Physics, University of California, 1156 High St., Santa Cruz, CA 95064, USA}\affiliation{Santa Cruz Institute for Particle Physics, Santa Cruz, CA 95064, USA}

\date{\today}
\begin{abstract}
\noindent We explore the phenomenology of a class of models where the dark matter particle can inelastically up-scatter to a heavier excited state via off-diagonal dipolar interactions with the interstellar plasma (gas or free electrons). The heavier particle then rapidly decays back to the dark matter particle plus a quasi-monochromatic photon. For the process to occur at appreciable rates, the mass splitting between the heavier state and the dark matter must be comparable to, or smaller than, the kinetic energy of particles in the plasma. As a result, the predicted photon line falls in the soft X-ray range, or, potentially, at arbitrarily lower energies. We explore experimental constraints from cosmology and particle physics, and present accurate calculations of the dark matter thermal relic density and of the flux of monochromatic X-rays from thermal plasma excitation. We find that the model provides a natural explanation for the observed 3.5 keV line from clusters of galaxies and from the Galactic center, and is consistent with null detections of the line from dwarf galaxies. The unique line shape, which will be resolved by future observations with the Hitomi (formerly Astro-H) satellite, and the predicted unique morphology and target-temperature dependence will enable easy discrimination of this class of models versus other scenarios for the generation of the 3.5 keV line or of any other unidentified line across the electromagnetic spectrum.
 \end{abstract}
%

\maketitle

\section{Introduction}
The particle nature of dark matter remains a mystery. Astronomical observations can be directly used to constrain or detect certain models of particle dark matter \cite{Feng:2014uja}. Dark matter pair-annihilation or decay generically produces photons, either promptly or through the decay of products of the annihilation or decay event; photons also arise from the secondary emission of the produced electrons and positrons \cite{Profumo:2010ya}. Other mechanisms producing electromagnetic emission from dark matter include, for instance, the Primakoff-like conversion of axion-like particles into photons in the presence of an external magnetic field \cite{Graham:2015ouw}.

Here, we present a completely novel mechanism to detect dark matter with astronomical observations. Our idea is that the dark matter sector consists of two physical states: a light state which is stable and is the dark matter particle, and a second, heavier state. The two dark-sector particles interact with themselves and with Standard Model particles through an effective operator which is an inelastic electric or magnetic dipole interaction term. This operator is responsible for two key features of our model:
\begin{enumerate}
\item the dark matter relic density, which is set almost entirely by co-annihilation processes;
\item the production of quasi-monochromatic photons, with typical energies corresponding to the mass splitting between the heavier state and the dark matter particle.
\end{enumerate}
The latter process produces a detectable flux of photons if the excitation rate for the heavy particle is large, and if the heavy particle decays quickly to the dark matter particle and a photon.

In this study we explore in detail the phenomenology and properties of this {\em novel indirect dark matter detection channel}. In particular, we show that the thermal relic density of the dark matter is easily accommodated for the same choice of parameters for which our model predicts a detectable flux of X-ray photons from excitations of the dark matter by collisions with the interstellar plasma, and for a broad range of masses. 

The signal strength predicted in our model depends on a peculiar combination of the dark matter number density times the interstellar plasma number density, which falls in the class of signal morphology  explored for example in Ref.~\cite{Carlson:2015daa}. The signal also depends on the kinetic energy of the plasma particles, and thus if the plasma is in thermal equilibrium, the plasma temperature is a key factor as to whether or not the excitation rate is significant. As a result, systems such as clusters of galaxies, which host abundant dark matter and thermal plasma with characteristic temperatures of a few keV, are expected to produce bright dark matter de-excitation lines. Also, we expect to detect this line in the Milky Way center, a relatively nearby location which again possesses both large plasma and dark matter densities. However, in our scenario no signal is to be expected from local dwarf galaxies, which have very small, if any, interstellar gas. Likewise, we do not expect any signal from small, distant galaxies.

Interestingly, the generic features expected in our model match observations of a recently discovered X-ray line at 3.5 keV, whose origin remains somewhat controversial. The line has been discovered in 2014 in observations of individual and stacked clusters of galaxies \cite{Bulbul:2014sua}. A line at the same energy was subsequently discovered in the center of the Milky Way \cite{Jeltema:2014qfa}, while its detection in M31 is debated and, at best, inconclusive \cite{Boyarsky:2014jta, Jeltema:2014qfa, Jeltema:2014mla}. On the other hand, no signal was detected neither in observations of dwarf spheroidal galaxies \cite{Malyshev:2014xqa}, which most notably includes recent, deep ($\sim1.6$ Msec) XMM observations of the Draco dwarf galaxy \cite{Jeltema:2015mee}, nor in stacked observations of galaxies and groups of galaxies \cite{Anderson:2014tza}.

Some excitement arose from the detection of the 3.5 keV line based on the statement in Ref.~\cite{Bulbul:2014sua} that the most plausible elemental line around 3.5 keV, from atomic de-excitation transitions of He-like potassium ions (K XVIII), would require an overabundance of K compared to solar of about 30, and thus would be ``physically difficult to understand''. Additionally, the energy and brightness of the line was found to be in principle compatible with what expected from the radiative decay of 7.1 keV sterile neutrinos with lifetimes of the order of $10^{29}$ s \cite{Bulbul:2014sua}. Other models have since then been extensively discussed in connection with a possible exotic origin of the line (see e.g. the recent review \cite{Adhikari:2016bei}; for models related to what we discuss here see in particular Ref.~\cite{Frandsen:2014lfa, Cline:2014eaa, Farzan:2014foo, Modak:2014vva, Finkbeiner:2014sja}).

There are convincing reasons to believe that the line does in fact originate from K XVIII transitions. The original argument in Bulbul et al \cite{Bulbul:2014sua} that this is unlikely because the required K overabundance compared to solar would be on the order of 30 is very likely incorrect for at least two reasons: 
\begin{enumerate}
\item The K solar abundance utilized in Ref.~\cite{Bulbul:2014sua} is the {\em photospheric} abundance, rather than the {\em coronal} abundance, which is about one order of magnitude larger \cite{Phillips:2015wla} and which is the relevant quantity as a proxy to the K abundance in the interstellar medium; 
\item The temperature models utilized in Ref.~\cite{Bulbul:2014sua} are skewed towards large temperatures (compared for example to what inferred from same-element ratios such as Ca XIX to Ca XX) resulting in a brightness for the K line suppressed by up to one order of magnitude (see e.g. fig.~4 in Ref.~\cite{Jeltema:2014mla}).
\end{enumerate}

Perhaps even more critically, Ref.~\cite{Carlson:2014lla} showed that the morphology of the 3.5 keV photons from Perseus and from the Milky Way matches closely the morphology of other elemental lines, rather than what expected from, for example, dark matter decay. This morphology by itself rules out a dark matter decay interpretation for the line \cite{Carlson:2014lla}. If an exotic origin is invoked, {\em the associated line emission should correlate spatially quite closely with the hot plasma in clusters and, possibly, in the Galaxy}. One such possibility is axion-like conversion in magnetic fields \cite{Cicoli:2014bfa}, although this scenario is not necessarily directly connected with dark matter, and the model parameters are tuned {\em ad hoc} to explain the observed signal.

Within the context of the scenario we consider here, it is instead natural to have a {\em thermal relic} dark matter candidate that produces a 3.5 keV line with (i) the required morphology, (ii) the required intensity to explain observations in clusters and in the Milky Way center, and that (iii) has suppressed emission from systems with low plasma temperatures and densities, such as dwarf galaxies. 

Our model is rather economical from the standpoint of input parameters. In fact, the model is entirely defined by (1) the masses of the two particles (or, equivalently, the dark matter particle mass and the mass splitting of the heavier state), (2) the effective electric and magnetic dipole moment couplings, $c_{E,M}/\Lambda$. As we will show below, the {\em magnetic} dipole controls the thermal relic density (the electric dipole featuring a $p$-wave suppression in the co-annihilation cross section), while the {\em electric} dipole dominates the scattering off of free electrons and protons in the interstellar plasma (the electrons dominating the rate at low dark matter masses,  $m_X\lesssim 50$ MeV, and the protons for larger dark matter masses, $m_X\gtrsim 50$ MeV).

The scenario we discuss here generically produces a bright, detectable X-ray (or lower energy) line, with the line width given by a geometric average of the dark matter and interstellar plasma velocity dispersions. Thus, as long as the line is resolved, for example with the expected energy resolution of the recently launched Hitomi (formerly known as Astro-H) satellite \cite{2014arXiv1412.1176K}, this scenario is observationally distinguishable and unique from both thermal plasma emission and from other new physics models. 

In connection with the question of the nature of the 3.5 keV line, our models explains the observational features of the line as a result of excitations generated by the scattering of the dark matter off of electrons and protons in the thermal plasma, as long as the plasma temperature is large enough to allow the excitation transition. The resulting morphology traces the product of the plasma density and the dark matter density, in qualitative agreement with what was observed in \Ref{Carlson:2014lla}. Additionally, the dark matter particle is naturally produced as a cold thermal relic from the early universe, dominantly from coannihilation processes. Crucially, we stress that the model we propose as a possible counterpart to the 3.5 keV line can be falsified with forthcoming observations with Hitomi, and would be strikingly different than for example axion-like particle conversion or dark matter decay.

We present the results of our study as follows. We introduce in Section~\ref{sec:EFT} the effective field theory (EFT) framework to investigate X-ray production from dark matter excitations. Such an EFT captures a large class of plausible UV completions, which we mention in what follows, and it allows a simple analysis in terms of two masses and of two coupling parameters. We identify the allowed range for these parameters in Section~\ref{sec:exp}, where we impose current experimental bounds. Dark matter production in the early universe through thermal freeze-out is discussed in Section~\ref{sec:DMprod}. With experimental and relic density constraints at hand, we finally compute the flux of X-rays from dark matter excitations and decays. General fluxes for arbitrary dark matter mass and mass splitting are presented in Section~\ref{sec:Xrays}, which ends with an analysis of the specific case of the 3.5 keV X-ray line. Conclusions are given in Section~\ref{sec:conclusions}.

\section{Effective Interactions for Inelastic Dark Matter}
\label{sec:EFT}

We introduce a simple EFT for the dark sector of this theory. We augment the Standard Model (SM) of particles physics with two additional gauge-singlet Weyl fermions $\xi$ and $\eta$, which are the only particles taken odd under a $\ZZ_2$ symmetry. As a consequence of this discrete symmetry, the particle corresponding to the lighter mass eigenstate is stable. The most general mass Lagrangian for the new degrees of freedom reads
\be
\mathcal{L}_{\rm mass} = - \mu \, \xi \eta - \frac{1}{2} \delta_\xi \, \xi \xi  - \frac{1}{2} \delta_\eta \, \eta \eta + {\rm h.c.} \ .
\label{eq:LagMass}
\ee
The EFT is valid only below a cutoff scale $\Lambda$, which is interpreted as the mass of some heavy particles we integrate out to generate the effective interactions between the SM and the new fermions. The mass parameters $(\mu, \delta_\xi, \delta_\eta)$ are consistently taken below the EFT cutoff. 

SM gauge invariance and the $\ZZ_2$ symmetry\footnote{The operators $L H \xi$ and $L H \eta$ would be allowed in the absence of the $\ZZ_2$ symmetry. Here, $L$ and $H$ are the SM lepton and Higgs doublets, respectively. These operators, together with the Majorana mass terms in \Eq{eq:LagMass}, would violate lepton number.} forbid any renormalizable interaction with SM fields. The lowest-order non-renormalizable interactions one can write down are the electric and magnetic dipole moments
\be
\begin{split}
\mathcal{L}_{\rm EFT} = & \,  - \frac{c_M}{2 \Lambda} \; \overline{\psi_D} \Sigma^{\mu\nu} \psi_D  \, F_{\mu\nu} + \\ & 
 - \frac{c_E}{2 \Lambda} \; \overline{\psi_D} \Sigma^{\mu\nu} \, i \gamma^5 \psi_D \, F_{\mu\nu} \ .
 \label{eq:EFT}
\end{split}
\ee
Here, we gather the two Weyl fermions $\xi$ and $\eta$ together to form a Dirac fermion $\psi_D$ as follows:
\be
\psi_D = \left( \begin{array}{c} \xi \\ \eta^\dag \end{array} \right) \ .
 \label{eq:DiracField}
\ee
We also define the antisymmetric tensor
\be
\frac{1}{2} \Sigma^{\mu\nu} = \frac{i}{4} \left[ \gamma^\mu , \gamma^\nu \right]  \ .
\ee
The operator with coefficient $c_M$ ($c_E$) is a CP-even(-odd) magnetic (electric) dipole moment. As we explicitly discuss in Section~\ref{sec:DMprod}, thermal freeze-out is likely to be dominated by the magnetic dipole interactions, since annihilations mediated by the electric dipole are $p$-wave suppressed. The situation is reversed for the dark matter (DM) excitations, and in Section~\ref{sec:Xrays} we show that interactions mediated by the electric dipole moment utterly dominate the up-scattering rate. 

The analysis of a microscopic origin for the effective interactions in \Eq{eq:EFT} is beyond the scope of this work, and UV-complete models can be constructed along the lines of e.g. \Ref{Lee:2014koa}.

Our model bears some similarity with the ``exciting dark matter'' (XDM) framework originally discussed in Ref.~\cite{Finkbeiner:2007kk}, and specialized to provide an explanation to the 3.5 keV line in Ref.~\cite{Finkbeiner:2014sja}. In the XDM setup there also exist two states, in fact two Weyl fermions like in our case, but interactions are mediated by a light vector mediator; for small mass splittings (below the electron pair threshold) the excited state is metastable on cosmological timescales, and in Ref.~\cite{Finkbeiner:2014sja} the authors then introduce an off-diagonal dipolar interaction for the purpose of having the decay happen with lifetimes shorter than the age of the universe.

There are several key differences between our setup and the XDM framework. First, in XDM the thermal relic density and the excitation process are both mediated by the light mediator, which is not present in our setup, where instead everything is accomplished through the dipole operator. The latter is typically at the electroweak scale in our setup, while it can be at a much higher scale in the XDM framework \cite{Finkbeiner:2014sja}. Second, the mechanism for excitation is the pair-annihilation of the light state in the heavier one. This mechanism dictates that the mass splitting be of the same order as the kinetic energy of the light state, and thus imposes certain requirements on the mass spectrum and on the average velocity which do not exist in our framework. As a result, the particle masses in the two models are at vastly different scales. Lastly, the predicted morphology and properties of the signal are entirely different: in XDM, the signal strength depends upon the line-of-sight dark matter density squared, with some requirement on the average velocity in the thermal average for the cross section; this is entirely different from our model, where the signal strength is given by the line-of-sight integral of the product of the dark matter times the plasma density; finally, the width of the line is also different in the two scenarios.

\subsection{Fermion Mass Spectrum}

The mass eigenstates for the new particles can be found by diagonalizing the fermion mass matrix 
\be
m_{\rm fermion} = \left( \begin{array}{cc} \delta_\xi & \mu  \\ \mu & \delta_\eta \end{array} \right) \ ,
\ee 
which follows from the Lagrangian in \Eq{eq:LagMass}. The three mass parameters are in general complex numbers. We always have the freedom to redefine the fields $\xi$ and $\eta$ to make two mass parameters real and positive. Here, we assume that all the mass parameters are real and positive, and the dipole operators in \Eq{eq:EFT} are given in the basis where this is the case. 

We find it convenient to introduce the parameter
\be
\epsilon \equiv \frac{\delta_\xi - \delta_\eta}{2 \mu} \ .
\ee
The exact mass eigenvalues can be expressed as follows
\begin{align}
m_1 = & \, \mu \sqrt{1 + \epsilon^2} - \frac{\delta_\xi + \delta_\eta}{2} \ ,  \\
m_2 = & \, \mu \sqrt{1 + \epsilon^2} + \frac{\delta_\xi + \delta_\eta}{2} \ .
\end{align}
We are ultimately interested in spectra where the mass splitting, of the order of few keV, is always much smaller than the overall mass scale for the new states. This allows us to express the mass eigenvalues in the $\epsilon \ll 1$ limit
\begin{align}
\label{eq:m1} m_1 \simeq & \, \mu - \frac{\delta_\xi + \delta_\eta}{2} \equiv m_\chi \ , \\
\label{eq:m2} m_2 \simeq & \, \mu + \frac{\delta_\xi + \delta_\eta}{2} \equiv m_\chi + \delta \ .
\end{align}
Here, we define $m_\chi$ to be the mass of the stable DM particle, and we denote the mass splitting with the excited state by $\delta$. Observationally, $\delta$ is a quantity of the utmost importance, as it sets the photon energy for the photon produced in the $\chi_2\to\chi_1+\gamma$ decay. The $\epsilon \ll 1$ limit, necessary in our framework to get a small relative mass splitting, can be justified by a hypothetical $U(1)$ symmetry in the UV complete theory that protects the Majorana mass terms $\delta_{\xi, \eta}$. Thus this limit can be quite natural. Strictly speaking, we do not need to forbid Majorana masses to be in this regime of validity, since all we need is the degeneracy between the two Majorana masses in order to have $\delta_\xi - \delta_\eta \ll \mu$. The ``line'' generated in the $\chi_2\to\chi_1+\gamma$ decay is thus at an energy that is effectively a free parameter in our scenario. 

The mass eigenstates can also be computed analytically. Here, we report the relevant expressions, once again in the $\epsilon \ll 1$ limit,
\begin{align}
\label{eq:chi1} \chi_1 = & \, \frac{i}{\sqrt{2}} \left( - \xi + \eta \right) \ , \\
\label{eq:chi2} \chi_2 = & \, \frac{1}{\sqrt{2}} \left( \xi + \eta \right) \ ,
\end{align}
corresponding to the mass values in \Eqs{eq:m1}{eq:m2}, respectively. 

\subsection{Interactions for Mass Eigenstates}

We conclude this Section with the effective interactions in \Eq{eq:EFT} for the mass eigenstates identified in \Eqs{eq:chi1}{eq:chi2}. We express the resulting Lagrangian in terms of the four-component Majorana fermions
\be
\psi_1 = \left( \begin{array}{c} \chi_1 \\ \chi_1^\dag \end{array} \right) \ , \qquad \qquad
\psi_2 = \left( \begin{array}{c} \chi_2 \\ \chi_2^\dag \end{array} \right) \ ,
 \label{eq:MajoranaFields}
\ee
and we find 
\be
\mathcal{L}_{\rm EFT} = - \frac{i}{2 \Lambda} \, \overline{\psi_2} \, \Sigma^{\mu\nu} \left( c_M + i \, c_E  \gamma^5 \right) \psi_1 \, F_{\mu\nu} \ .
\label{eq:EFT2}
\ee
It is straightforward to use the properties of the four-component Majorana spinors in \Eq{eq:MajoranaFields} to check that this Lagrangian is hermitian. Whenever the mass splitting plays a crucial role, as for example in the calculation of the excited state lifetime or the up-scattering rate, we use the interactions as given in \Eq{eq:EFT2}. However, to compute the thermal relic density or limits from virtual DM effects, the effect of the mass splitting is completely irrelevant. In these latter cases we perform our calculations in what we call the ``Dirac limit", namely when the interactions can be taken as in \Eq{eq:EFT}.

\section{Experimental Constraints}
\label{sec:exp}

In this Section we analyze what region of the EFT parameter space is allowed by current experimental bounds. We then study, in the allowed range of parameters, the thermal production of DM and the predicted flux of X-rays from DM excitation in Sec.~\ref{sec:DMprod} and \ref{sec:Xrays}, respectively. We do not report here constraints not relevant to our analysis, such as DM-induced contributions to: muon anomalous magnetic moment, electric dipole moments of charged SM fermions, $Z$-pole observables, invisible $B$ and $K$ meson decays. We checked that all of those constraints are not competitive with the one coming from the electromagnetic coupling running we discuss below~\cite{Sigurdson:2004zp}. Additionally, searches  for mono-photon and mono-jet events at colliders are performed at energy scales above the typical cutoff values we are interested in ($\Lambda \simeq 200 \, {\rm GeV}$). A correct interpretation of these negative searches would thus require the specification of the underlying UV-complete theory giving the effective interactions in \Eq{eq:EFT}, which is model-dependent and beyond the scope of this work.

\subsection{Big Bang Nucleosynthesis} 
\begin{figure}
\begin{center}
\includegraphics[width=0.4\textwidth]{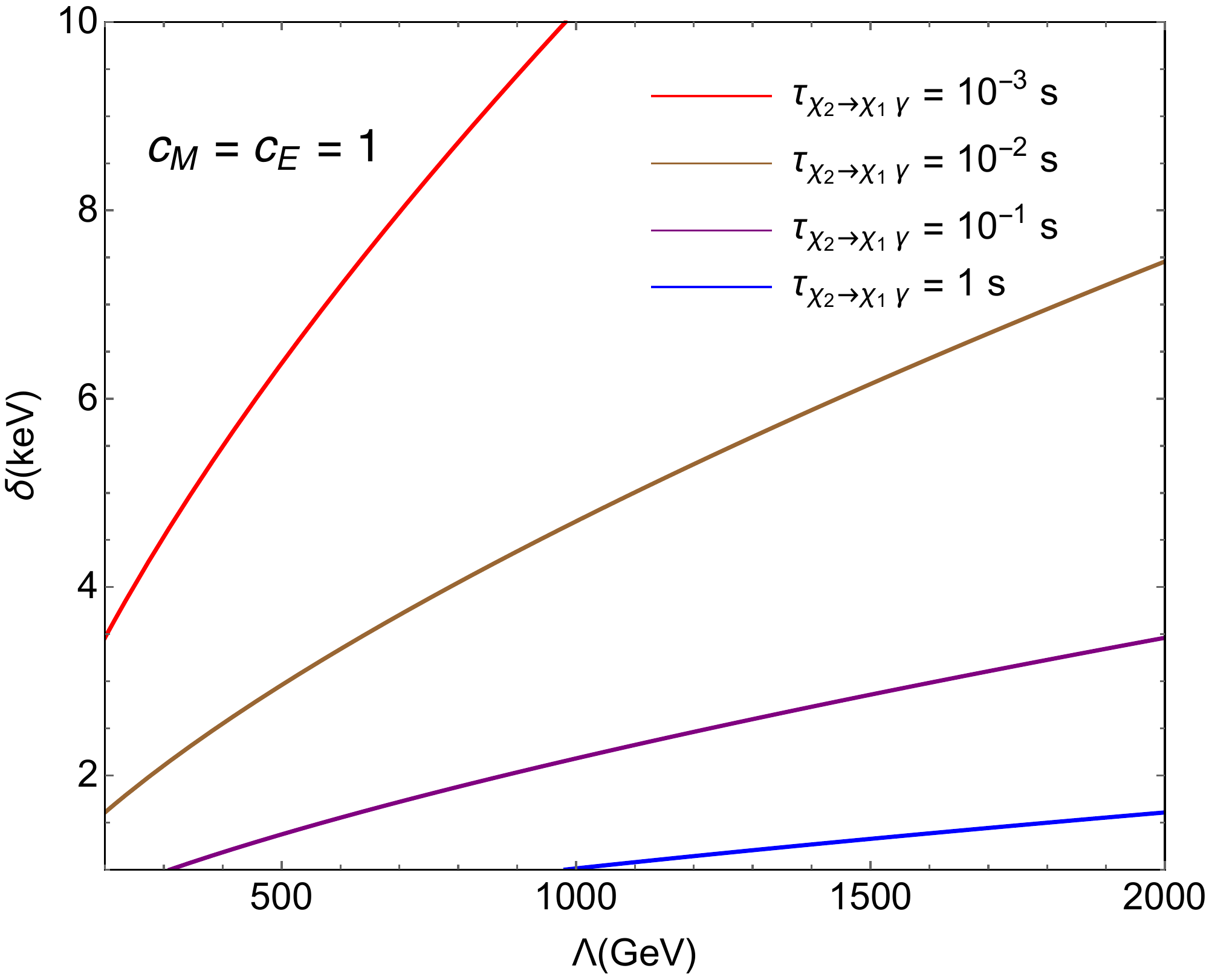}  
\end{center}
\caption{Lifetime of the excited state decaying to photon through the reaction $\chi_2 \, \rightarrow \, \chi_1 \gamma$. We plot iso-contours in the $(\Lambda, \delta)$ correspondent to different lifetimes. We fix the dipole Wilson coefficients to $c_M = c_E = 1$, for different values the lifetime can be obtained by simply rescaling as in \Eq{eq:lifetimenum}.} 
\label{fig:tau}
\end{figure}

Light degrees of freedom can alter Big Bang Nucleosynthesis (BBN) and spoil the successful prediction of light elements abundance. We impose two types of BBN constraints:
\begin{enumerate}
\item We consider the possibility for DM to freeze-out {\em after} neutrino decoupling. DM annihilation would then heat the SM plasma with respect to the cosmic neutrino background, {\em decreasing} the value of the effective relativistic number of degrees of freedom at the BBN epoch $N_{\rm eff}$~\cite{Nollett:2013pwa}. We impose the conservative bound of the DM mass $m_\chi \gtrsim 10 \, {\rm MeV}$, such that DM freeze-out takes place always before neutrinos decouple.
\item An additional concern pertains the timing for the decay of the excited states $\chi_2$. As will be explained in detail in Sec.~\ref{sec:DMprod}, thermal freeze-out democratically populates the universe with $\chi_2$ and $\chi_1$ through co-annihilations. The excited state $\chi_2$ has a decay width
\begin{align}
\Gamma_{\chi_2 \rightarrow \chi_1 \gamma} &=  \frac{c_M^2 + c_E^2}{8 \pi \Lambda^2} \frac{(m_2^2 - m_1^2)^3}{m_2^3} \nonumber \\
&\simeq \frac{c_M^2 + c_E^2}{\pi \Lambda^2} \, \delta^3 \ .
\end{align}
The first expression is general and does not contain any assumption about the relative size of the mass splitting. The last expression gives the decay width in the small mass splitting limit, where the masses are given by \Eqs{eq:m1}{eq:m2}. Interestingly, in such a limit the decay width is controlled only by the mass splitting $\delta$ and not by the overall mass scale $m_\chi$. Plugging in typical values for the parameters we are interested in, we find the lifetime $\tau_{\chi_2 \rightarrow \chi_1 \gamma} = \Gamma^{-1}_{\chi_2 \rightarrow \chi_1 \gamma} $ to be typically shorter than the BBN timescale:
\be
\begin{split}
 \tau_{\chi_2 \rightarrow \chi_1 \gamma}   = & \,   9.7 \times 10^{-4} \, {\rm s} \;  \left(\frac{2}{c_M^2 + c_E^2}\right) 
  \\ &  \, \times  \left( \frac{3.5 \, {\rm keV}}{\delta} \right)^3 \left( \frac{\Lambda}{200 \, {\rm GeV}} \right)^2  \ .
\end{split}
\label{eq:lifetimenum}
\ee
A thorough exploration of the $(\Lambda, \delta)$ plane is provided in Fig.~\ref{fig:tau}, where we show iso-contours of the excited state lifetime (the scaling for different values of $c_M$ or $c_E$ is entirely trivial). For the parameter range we are interested in this is always below $\tau_{\rm BBN} \simeq 1 \, {\rm s}$ and BBN is thus safe.
\end{enumerate}

\subsection{Cosmic Microwave Background}

Out-of-equilibrium DM annihilation after freeze-out can change the recombination history and leave an imprint in the CMB spectrum~\cite{Galli:2009zc,Slatyer:2009yq,2012PhRvD..85d3522F}.  CMB anisotropies bound the DM annihilation strength $\langle \sigma v_{{\rm rel}}\rangle$, putting an upper limit on the quantity
\be
p_{\rm ann} = f_{\rm eff} \, \frac{\langle \sigma v_{{\rm rel}}\rangle}{m_{\rm DM}} \ .
\ee
Here, the efficiency parameter $f_{\rm eff}$ depends on the specific DM annihilation channel. The energy injection takes place over a narrow window of redshift values, thus it is reasonable to approximate $f_{\rm eff} \simeq {\rm const}$~\cite{2011PhRvD..84b7302G,Giesen:2012rp,2012PhRvD..85d3522F}. The values for $f_{\rm eff}$ as a function of the DM mass and for different annihilation channels can be found in \Ref{Slatyer:2015jla}. The latest Planck results on CMB anisotropies~\cite{Ade:2015xua} give the limit
\be
\langle \sigma v_{{\rm rel}}\rangle <  f_{\rm eff} \, \times \, 4.1 \times 10^{-28} \, {\rm cm}^3 {\rm s}^{-1} \, \left(\frac{m_{\rm DM}}{{\rm GeV}} \right) \ .
\ee
For $m_{\rm DM}$ around $1 \, {\rm GeV}$, this bound is approximately two orders of magnitude below the value needed at freeze-out $\langle \sigma v_{{\rm rel}}\rangle_{\rm th} = 3 \, \times \, 10^{-26} \, {\rm cm}^3 {\rm s}^{-1}$ in order to reproduce the observed DM density.

\begin{figure}
\begin{center}
\includegraphics[width=0.49\textwidth]{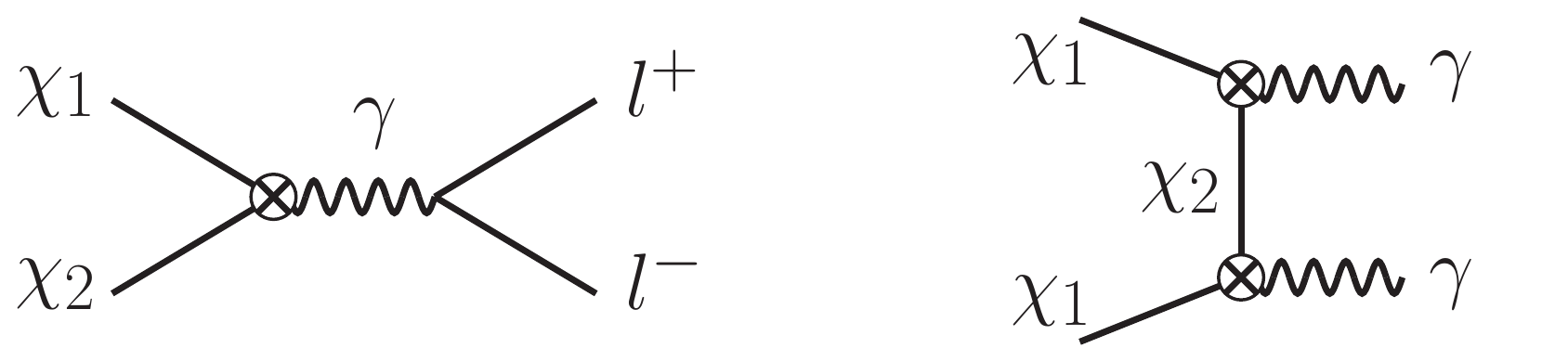}  
\end{center}
\caption{Feynman diagrams for DM (co)-annihilations. The co-annihilations to leptons (left), as well as the ones to hadrons computed as described in the text, totally dominate the $\Lambda^{-4}$ suppressed annihilation to photons (right). In both cases, the crossed
circle denotes the insertion of the effective magnetic or electric dipole operator.} 
\label{fig:FeynDiag}
\end{figure}

We have three possible DM annihilation final states: leptons, hadrons and photons. The associated Feynman diagrams are shown in Fig.~\ref{fig:FeynDiag}. Detailed results for the cross sections can be found in Sec.~\ref{sec:DMprod}, where we discuss thermal freeze-out. For the purpose of this CMB bound discussion, we limit ourselves to two observations: 
\begin{enumerate}
\item The annihilation to leptons and hadrons proceeds through an $s$-channel photon exchange with an off-diagonal vertex. At the time of recombination all the $\chi_2$ have decayed to the stable $\chi_1$, therefore this contribution is absent. 
\item Annihilation to two photons, through the diagram sketched on the right of Fig.~\ref{fig:FeynDiag}, is still possible, albeit this channel is suppressed by a double insertion of the dimension-5 dipole operator. This channel has minimal impact on the thermal production of dark matter. As explicitly shown in \Eq{eq:sigmagammagamma} in the next Section, this $m_\chi^2 / \Lambda^2$ suppression is quite severe and makes this contribution irrelevant for CMB constraints. 
\end{enumerate}
We conclude that CMB limits do not constrain our framework. 

\subsection{Direct Detection}

The long-range dipole interactions are responsible for quite sizable rates in direct detection (DD) experiments. As is well known~\cite{Sigurdson:2004zp,Barger:2010gv,Banks:2010eh}, this is one of the main reasons why the Wilson coefficients of the dipole operators are severely constrained. However, in our framework as outlined in Sec.~\ref{sec:EFT} the dipole interactions are off-diagonal and DD can only go through an inelastic channel~\cite{TuckerSmith:2001hy}. Hence DD limits depend on the specific value of the splitting $\delta$, controlling  the threshold velocity for the inelastic scattering, and also on the specific experiment because of different energy thresholds to detect the recoil. 

We focus our analysis on a mass range $m_\chi \lesssim 1 \, {\rm GeV}$, where the nuclear recoil energy for DM scattering off of nuclei is way below any current experimental threshold, even for the case of elastic scattering. In such a low DM mass range the most promising probe for DD is scattering off of electrons. Xenon10 already put limits on this sub-GeV DM mass range~\cite{Essig:2011nj,Essig:2012yx}, which will be further improved by future experiments~\cite{Graham:2012su,Essig:2015cda}. However, these bounds consider the elastic case. The condition to have inelastic scattering reads
\be
\delta \; < \; \frac{1}{2} \frac{m_\chi m_e}{m_\chi + m_e} \, v^2 \simeq m_e v^2,
\ee
where $\delta$ as above is the mass splitting between the dark sector fermions. While $\delta$ is an arbitrary parameter, we are interested in keV emission, thus $\delta\simeq$ keV. For a typical WIMP velocity $v \simeq 10^{-3}$ this condition only applies to very small mass splitting, on the order of fractions of an eV. Given the mass splitting range of a few keV we are interested in there is no signal in this type of experimental set-up. Thus as long as $m_\chi \lesssim 1 \, {\rm GeV}$ our model is not constrained by DD exclusion bounds. 

\subsection{Electromagnetic Coupling Running}

The contact interaction between DM particles and the photon alters the running of the electromagnetic coupling at low energies. This effect can be quantified by computing the one-loop contribution to the photon self-energy accounting for DM particles in the loop. We defer the details of the calculation to App.~\ref{app:running}.

The running of the electromagnetic coupling is affected as follows:
\be
\alpha_{\rm em}(q^2) = \frac{\alpha_0}{1 - C \, \Delta \alpha(q^2)} \ .
\ee
The fine structure constant $\alpha_0$ is measured with extremely high precision via the anomalous electron magnetic moment~\cite{Mohr:2015ccw}. The function $\Delta \alpha(q^2)$ accounts for only the SM degrees of freedom~\cite{Steinhauser:1998rq,Cabibbo:1961sz,Burkhardt:2001xp,Burkhardt:2011ur}, therefore in the absence of the DM particles we would just have $C = 1$. The DM contribution enters $C$ as follows
\be
C = 1 + \frac{\Pi^{\rm ren}_{\rm dip}(q^2)}{\Delta \alpha(q^2)} \ ,
\label{eq:Cdipole}
\ee
where the renormalized photon self-energy $\Pi^{\rm ren}_{\rm dip}(q^2)$ from virtual DM particles is computed in App.~\ref{app:running}. Its final expression is given in \Eq{eq:PiRenDip2}. We take the the LEP measurement of the running electromagnetic coupling~\cite{Acciarri:2000rx,Achard:2005it} that are all performed at momentum transfers larger than the DM masses under consideration. For our purposes we can simplify \Eq{eq:PiRenDip2} in this limit. We find
\be
\begin{split}
C - 1 = & \, \frac{(- q^2)}{\Delta \alpha(q^2)} \frac{c_M^2 + c_E^2}{4 \pi^2 \Lambda^2}   \int_0^1 dx \; x (1 - x) \, \times \\ & \ln\left(\frac{m_\chi^2}{m_\chi^2 + (- q^2) x (1 - x)}\right) \ .
\end{split}
\ee
Bhabha-scattering probes the electromagnetic coupling at space-like momentum transfer ($ - q^2 > 0$), therefore the above expression is always negative. Consequently, the dipole contribution to $C$, as parameterized in \Eq{eq:Cdipole}, is negative and, in our model, we always have $C \leq 1$.

We are ready to compare our theory prediction with LEP data. The analysis in \Ref{Achard:2005it} found the bound
\be
C = 1.05 \; \pm \; 0.07_{\rm stat} \, \pm \, 0.14_{\rm syst} \ .
\label{eq:C}
\ee
We evaluate the DM contribution for $m_\chi = 10 \, {\rm MeV}$, corresponding to the minimum DM mass we consider, which is also the case when the effect on the running is maximized. Plugging the lowest value of $(-q^2)$ within the range probed by the analysis in \Ref{Achard:2005it}, and imposing that we do not violate the bound in \Eq{eq:C} beyond $1 \sigma$, we find the following constraints of the combination of Wilson coefficients and the suppression scale
\be
\left( c_M^2 + c_E^2 \right) \left(\frac{200 \, {\rm GeV}}{\Lambda}\right)^2 \lesssim 1 \ .
\ee
For order one Wilson coefficients the running of the electromagnetic coupling requires $\Lambda \gtrsim 200 \, {\rm GeV}$. The rescaling for different Wilson coefficients is straightforward.

\section{Dark Matter Relic Density}
\label{sec:DMprod}

In this section we demonstrate that in our model, and for parameter values compatible with the constraints discussed above, the stable DM particle $\chi_1$ can be produced in the early universe through thermal freeze-out. As we have extensively discussed in the previous Section, we are interested in the DM mass range $m_\chi \gtrsim 10 \, {\rm MeV}$ and a mass splitting $\delta \simeq {\rm few} \, {\rm keV}$. Thus at the freeze-out epoch, happening when the universe had a temperature approximately $T_f \simeq m_\chi / 20$, we expect the excited state $\chi_2$ to be thermally populated; as a result, co-annihilations have to be accounted for. However, since we work in the $\delta \ll m_\chi$ regime, we can treat the $(\chi_1, \chi_2)$ system as a Dirac fermion with mass $m_\chi$ and compute the annihilation cross sections in the ``Dirac limit" of Sec.~\ref{sec:EFT}, namely by using the effective Lagrangian in \Eq{eq:EFT}.

Three possible (co-)annihilation channels keep the DM in thermal equilibrium at early times. Both electric and magnetic dipoles allow the DM to annihilate to lepton pair final states, as shown in the Feynman diagram on the left of Fig.~\ref{fig:FeynDiag}. In the non-relativistic limit, appropriate for a cold relic as in our case, we calculate a cross section
\be
\begin{split}
\sigma_{\chi \chi \; \rightarrow \; l^+ l^-} \, v_r \simeq & \,  \frac{\alpha_{\rm em}}{\Lambda^2} \, \left[c_M^2 + c_E^2 \frac{v_r^2}{12} \right]  \, \times \\ &   \left( 1 - \frac{m_l^2}{m_\chi^2} \right)^{1/2} \, \left( 1 +  \frac{m_l^2}{2 m_\chi^2}   \right) \ .
\end{split}
\label{eq:anntofexp}
\ee
As clear from the formula above, annihilation processes mediated by magnetic and electric dipole moments are $s$- and $p$-wave processes, respectively. For DM masses above the pion mass, the same interaction vertex with the photon is responsible for annihilation to hadrons. We evaluate this contribution by using the measured value of the observable
 \be
R_h(\sqrt{s}) = \frac{\sigma_{e^+ e^- \, \rightarrow \, {\rm hadrons}}}{\sigma_{e^+ e^- \, \rightarrow \, \mu^+ \mu^-}} \ .
\ee
We import numerical values for $R_h(\sqrt{s})$ from the Particle Data Group public webpage~\footnote{\href{http://pdg.lbl.gov/current/xsect/}{http://pdg.lbl.gov/current/xsect/}}, which gives the value of this observable for $\sqrt{s} > 0.36 \, {\rm GeV}$. We fill the gap in the region  $2 m_\pi \leq \sqrt{s} \leq  0.36 \, {\rm GeV}$ by using $e^+ e^- \, \rightarrow \pi \pi$ scattering data from \Ref{Ezhela:2003pp}. The annihilation cross section to hadrons results in
\be
\sigma_{\chi \chi \; \rightarrow \; {\rm hadrons}} = \mathcal{R}_h(\sqrt{s} = 2 m_\chi) \, \times \,  \sigma_{\chi \chi \; \rightarrow \; \mu^+ \mu^-}  \ .
\label{eq:anntohadrons}
\ee
A double insertion of dipole operators gives the annihilation to photons shown on the right of Fig.~\ref{fig:FeynDiag}. The resulting cross section 
\be
\sigma_{\chi \chi \; \rightarrow \; \gamma \gamma} \, v_r \simeq   \frac{(c_M^2 + c_E^2)^2}{4 \pi \, \Lambda^4} \, m_\chi^2 \ ,
\label{eq:sigmagammagamma}
\ee
suppressed by the fourth inverse power of $\Lambda$, is a sub-dominant contribution to the total annihilation cross section and it does not play any role at the freeze-out epoch.

The DM number density evolution is described by the Boltzmann equation
\be
\frac{d n_\chi}{d t} + 3 H n_\chi = -\langle \sigma v_{{\rm rel}}\rangle \left(n_\chi^2 - n_\chi^{{\rm eq}\,2}\right) \ .
\label{eq:BoltzFO}
\ee
The thermal average of the cross section for annihilation to leptons in \Eq{eq:anntofexp} results in
\be
\begin{split}
\langle\sigma_{\chi \chi \; \rightarrow \; l^+ l^-} \, v_r\rangle (T) \simeq & \,  \frac{\alpha_{\rm em}}{\Lambda^2} \, \left[c_M^2 + c_E^2 \frac{T}{2 \, m_\chi} \right]  \, \times \\ &   \left( 1 - \frac{m_l^2}{m_\chi^2} \right)^{1/2} \, \left( 1 +  \frac{m_l^2}{2 m_\chi^2}   \right) \ ,
\end{split}
\label{eq:anntofexpTH}
\ee
where $T$ is the temperature of the relativistic bath in thermal equilibrium. The total annihilation cross section includes three contributions. Processes with final state electrons, with cross section as in \Eq{eq:anntofexpTH}, are always kinematically allowed for the DM mass range we are interested in. Annihilations to muons, with cross section given in \Eq{eq:anntofexpTH}, and to hadrons, with cross section obtained through $R_h(\sqrt{s}) $ as in \Eq{eq:anntohadrons}, have to be accounted for only if kinematically allowed. 

The Boltzmann equation is conveniently solved in terms of the comoving number density $Y_\chi = n_\chi / s$, where $s$ is the entropy density of the relativistic species. Since we are interested in sub-GeV DM, thermal freeze-out is likely to occur during the QCD phase transitions. We take the QCD equation of state, necessary to evaluate the entropy density $s$ and therefore $Y_\chi$, from \Ref{Laine:2006cp}. At temperatures much lower than the one at the freeze-out epoch $T_f$, the comoving density approaches a constant value $Y^\infty_\chi = Y_\chi(T \gg T_f)$. The number and mass density at the present epoch are
\begin{align}
\label{eq:ninfty} n^\infty_\chi  = & \, 2 \times Y^\infty_\chi s_0  \ , \\
\rho^\infty_\chi  = & \,  m_\chi n^\infty_\chi \ ,
\end{align}
where we have for the current entropy density~\cite{Agashe:2014kda} 
\be
 s_0 = 2891.2 \, {\rm cm}^{-3} \ .
\ee
The factor of $2$ in \Eq{eq:ninfty} accounts for the fact that we are dealing with a Dirac fermion. Finally, we compute the dark matter contribution to the $\Omega$ parameter
\be
\Omega_\chi h^2 = \frac{\rho_\chi}{\rho_{{\rm cr}} / h^2} \ ,
\ee
where the critical density is given by~\cite{Agashe:2014kda} 
\be
\rho_{\rm cr} / h^2 = 1.05375 \times 10^{-5} \,  \GeV \, {\rm cm}^{-3} \ .
\ee

\begin{figure}
\begin{center}
\includegraphics[width=0.45\textwidth]{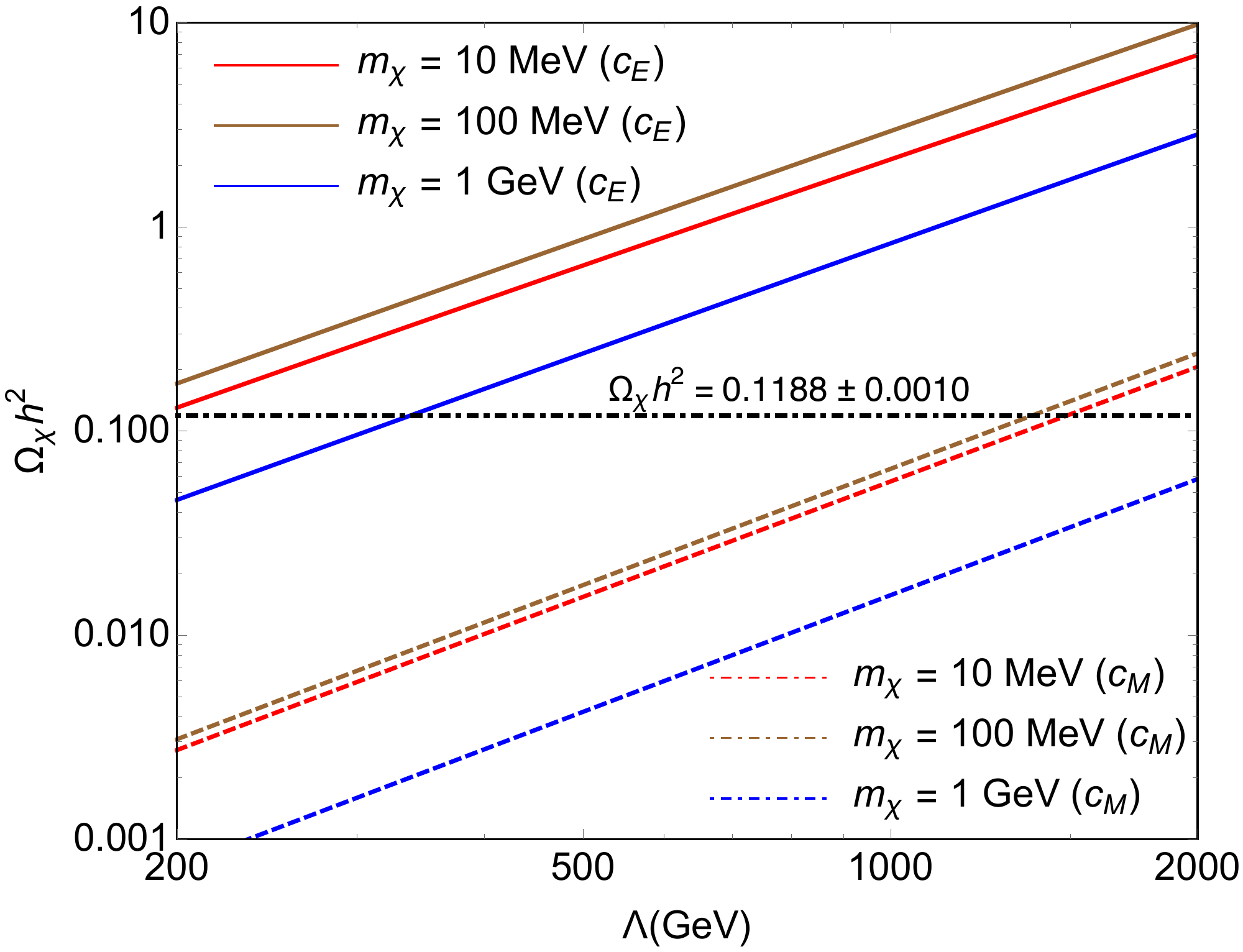}  
\end{center}
\caption{DM relic density as a function of $\Lambda$ for three different DM masses (lines with different colors) for purely magnetic dipole interactions ($c_M = 1$, dot-dashed lines) and purely electric dipole interactions ($c_E = 1$, solid lines).} 
\label{fig:relic}
\end{figure}

The output of this calculation has to confront the latest measured value by Planck $\Omega_{\rm DM} h^2 = 0.1188 \pm 0.0010$~\cite{Ade:2015xua}. Our final results are summarized in Fig.~\ref{fig:relic}, where we plot the relic density as a function of the suppression scale $\Lambda$. We choose three representative value for the DM mass, $m_\chi = (10 \, {\rm MeV}, 100 \, {\rm MeV}, 1 \, {\rm GeV})$, and for each case we compute the relic density for magnetic dipole ($c_M = 1$, dot-dashed lines) and electric dipole ($c_E = 1$, solid lines) interactions. While the figure was produced for $c_{M,E} = 1$, it is straightforward to obtain the relic density for arbitrary Wilson coefficients just by dividing the result of Fig.~\ref{fig:relic} by $c_{M,E}^{\,2}$.

A noteworthy feature of Fig.~\ref{fig:relic} is that the magnetic dipole lines are always well below the ones for electric dipole moments. This can be understood by looking at \Eq{eq:anntofexp}, which shows that the latter are $p$-wave processes and therefore yield a larger relic density. 

It is interesting to discuss the dependence on the DM mass. The annihilation cross section for $m_\chi = 10 \, {\rm MeV}$ and $m_\chi = 100 \, {\rm MeV}$ is dominated by electron/positron final states and it is approximately the same for both mass values. Despite the two cross sections being identical, the resulting DM relic density is not the same. This residual DM mass dependence is a consequence of the $g_*$ dependence on the freeze-out temperature $T_f \simeq m_\chi / 20$, one order of magnitude different in the two cases. The quite different DM relic density for $m_\chi = 1 \, {\rm GeV}$ is perhaps more obvious, as hadronic channels are kinematically available, suppressing the total relic density.

To summarize, both magnetic and electric dipole moments can reproduce the observed DM density. For each choice of the DM mass and the suppression scale $\Lambda$, all we have to do is to choose $c_{M}$ or $c_{E}$ such that $\Omega_\chi h^2 \simeq 0.12$. The relic density for arbitrary values of $c_{M,E}$ is obtained by taking the results shown in Fig.~\ref{fig:relic} and dividing the predicted value of the relic density in the figure by $c_{M,E}^{\,2}$. As we are about to see in the next Section the excitation rate is dominated by $c_E$, therefore once both couplings are present and comparable we have a remarkable feature of our model: The relic density and the X-ray lines rate are independently controlled by the magnetic and by the electric dipole moments, respectively. For the range of parameters we are interested in, Wilson coefficients for the magnetic dipole in the range $0.1 \lesssim c_M \lesssim 1$ are what is needed to account for the observed DM density.

\section{X-rays from dark matter excitation}
\label{sec:Xrays}

\begin{figure}
\begin{center}
\includegraphics[width=0.25\textwidth]{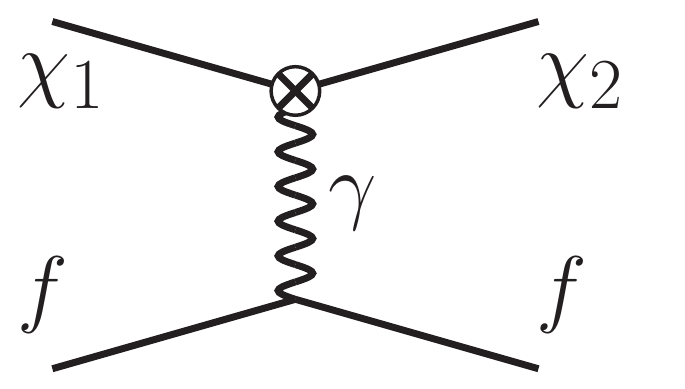}  
\end{center}
\caption{Feynman diagram for DM inelastic up-scattering. Here, $f$ can be either an electron or a proton. The crossed circle denotes the insertion of the effective magnetic or electric dipole operator.} 
\label{fig:FeynDiag2}
\end{figure}

In this Section we evaluate the predicted flux of X-ray photons originating from the excitation process
\be
\chi_1 \, f \; \rightarrow \; \chi_2 \, f \ ,
\label{eq:excitationreaction}
\ee
followed by the decay process
\be
\chi_2 \; \rightarrow \; \chi_1 \gamma \ .
\ee
The Feynman diagram for the up-scattering is shown in Fig.~\ref{fig:FeynDiag2}. The particle $f$ is a SM fermion present in the plasma. For simplicity we consider contributions from electrons and protons and neglect that from heavier elements. The excited state $\chi_2$ is quite short-lived compared to cosmological timescales, as explicitly shown in \Eq{eq:lifetimenum}. Hence once the DM up-scatters off of a plasma fermion into the $\chi_2$ particle, the subsequent decay back to the stable DM particle $\chi_1$ and the consequent emission of a X-ray photon are effectively instantaneous. The final state photon energy is equal to the mass splitting between the two fermion states (up to corrections due to the velocity distribution of the plasma fermion and of the dark matter, as discussed below):
\be
E_\gamma = \frac{m_{\chi_2}^2 - m_{\chi_1}^2 }{2 m_{\chi_2}} \simeq \delta \ .
\ee

The predicted X-ray flux in our model resulting from DM excitation and decay reads
\be
\Phi = \kappa_{\rm eff} \vev{\sigma v_{\rm rel}}  \ ,
\label{eq:PhiDef}
\ee
where $\kappa_{\rm eff}$ is the integral along the line of sight and over the appropriate (solid) angular region of interest $\Delta\Omega$ (corresponding to angles of aperture of around 6 arcmin for X-ray observations, unless otherwise specified) of the product of the plasma particles' number density $n_f(\vec r)$ times the dark matter number density $n_{\rm DM}=\rho_{\rm DM}(r)/m_{\chi}$, and has units of cm$^{-5}$,
\be
\kappa_{\rm eff} \equiv \int_{\Delta\Omega} d\Omega\int_{\rm l.o.s.} dl\ n_f(\vec r(l,\Omega)) \, n_{\rm DM}(\vec r(l,\Omega)),
\label{eq:neffdef}
\ee
where $f$ is either a proton or an electron. The thermally averaged cross section is defined as
\be
\vev{\sigma v_{\rm rel}} = \int d^3 v_\chi d^3 v_f  \; f_{\rm DM}(v_\chi) f_f(v_f)  \; \sigma v_{\rm rel} \ ,
\label{eq:sigmavdef}
\ee
where $f_{\rm DM}$ and $f_f$ are the normalized phase space distribution functions of the DM particle and the SM fermion, respectively. We consider for our analysis normalized Gaussian distributions
\be
f_i(v_i) = \frac{a_i^{3/2}}{\pi^{3/2}} e^{- a_i v_i^2} \ .
\label{eq:fvi}
\ee
where $i = \chi, f$. As derived in App.~\ref{app:sigmav}, the thermal average can be expressed in a very simple form, namely 
\be
\vev{\sigma v_{\rm rel}} =  \int^\infty_{v_{\rm rel}^{\rm min}} d v_{\rm rel} \; \sigma(v_{\rm rel}) \; \mathcal{F}(v_{\rm rel}).
\label{eq:sigmavrel}
\ee
The integration is over the relative velocity in the CM frame of the inelastic scattering in \Eq{eq:excitationreaction}. Such a collision can only take place for relative velocity values above the kinematic threshold
\be
v_{\rm rel}^{\rm min} = \left(\frac{m_\chi + m_f}{m_\chi m_f} \, 2 \delta \right)^{1/2}.
\label{eq:vmin}
\ee
The integrand in \Eq{eq:sigmavrel} is the product of the total inelastic cross section and a ``Kernel function''
\be
\mathcal{F}(v_{\rm rel}) = \frac{4  v_{\rm rel}^3}{\pi^{1/2}} \, \left(\frac{a_\chi a_f}{a_\chi + a_f}\right)^{3/2} 
\exp\left( - \frac{a_\chi a_f}{a_\chi + a_f} \, v_{\rm rel}^2 \right) \ .
\label{eq:Fvrel}
\ee
Here, the coefficients $a_{\chi, f}$ denotes the width of the thermal distribution in \Eq{eq:fvi}. In particular, for the two cases we are interested in we have, in natural units,
\begin{align}
a_\chi = & \, \frac{1}{v_0^2} \simeq 10^6 \ , \\
a_f = & \, \frac{m_f}{2 T_f} \ .
\end{align}

\subsection{General Features of the Rate}
\label{sec:ratesgeneral}

Before discussing the two cases we are interested in, namely the Perseus cluster of galaxies and the Galactic Center, we point out a few general facts about the X-ray line rates in our framework. The above discussion is very general and does not assume any specific type of interaction mediating the inelastic collision in \Eq{eq:excitationreaction}. We specialize now on the dipole interactions introduced in Section~\ref{sec:EFT}. We are considering an inelastic collision, therefore the cross section has to be computed by using the Lagrangian for mass eigenstates given in \Eq{eq:EFT2}.

The expression for the up-scattering cross section is quite involved. It is helpful to look at the limiting case when the mass splitting is negligible~\cite{Sigurdson:2004zp}:
\begin{align}
\label{eq:upscatt1} \frac{d \sigma}{d \Omega} = & \, \frac{c_M^2 \, e^2}{16 \pi^2 \Lambda^2} \;
\frac{1 + \frac{m_\chi (m_\chi - 2 m_f)}{(m_\chi + m_f)^2} \sin^2(\theta/2)   }{\sin^2(\theta/2)}   \ , \\
\label{eq:upscatt2} \frac{d \sigma}{d \Omega} = & \, \frac{c_E^2 \, e^2}{16 \pi^2 \Lambda^2} \;
\frac{1}{v_{\rm rel}^2 \, \sin^2(\theta/2)}  \ ,
\end{align}
for magnetic and electric dipole moments, respectively. Both expressions are divergent at small $\theta$. Furthermore, the cross section for the process mediated by the dipole is also divergent at small relative velocities. The introduction of a finite mass splitting regularizes all the above divergences and our final results are thus finite. The usefulness of Eqs.(\ref{eq:upscatt1}) and (\ref{eq:upscatt2}) is in the small velocity behavior: since we are interested in the non-relativistic regime, we expect the electric dipole contribution to dominate the total excitation rate.

This can be quantified by performing the full computation of the cross section and evaluating the thermal average as prescribed by \Eq{eq:sigmavrel}. Giving the scaling of our dipole interactions in \Eq{eq:EFT2}, the thermal average can be parameterized as follows:
\be
\vev{\sigma v_{\rm rel}} = \frac{e^2}{\Lambda^2} \left[ c_M^2 \, \Sigma_M(m_\chi, \delta) + c_E^2 \, \Sigma_R(m_\chi, \delta) \right] \ .
\label{eq:SigmaDef}
\ee
In this expression we assume the other parameters (i.e.\ the DM dispersion velocity, fermion mass and temperature) to be fixed, and therefore the functions $\Sigma_{M,E}$ to only depend on the DM mass and on the mass splitting. 

\begin{figure}
\begin{center}
\includegraphics[width=0.45\textwidth]{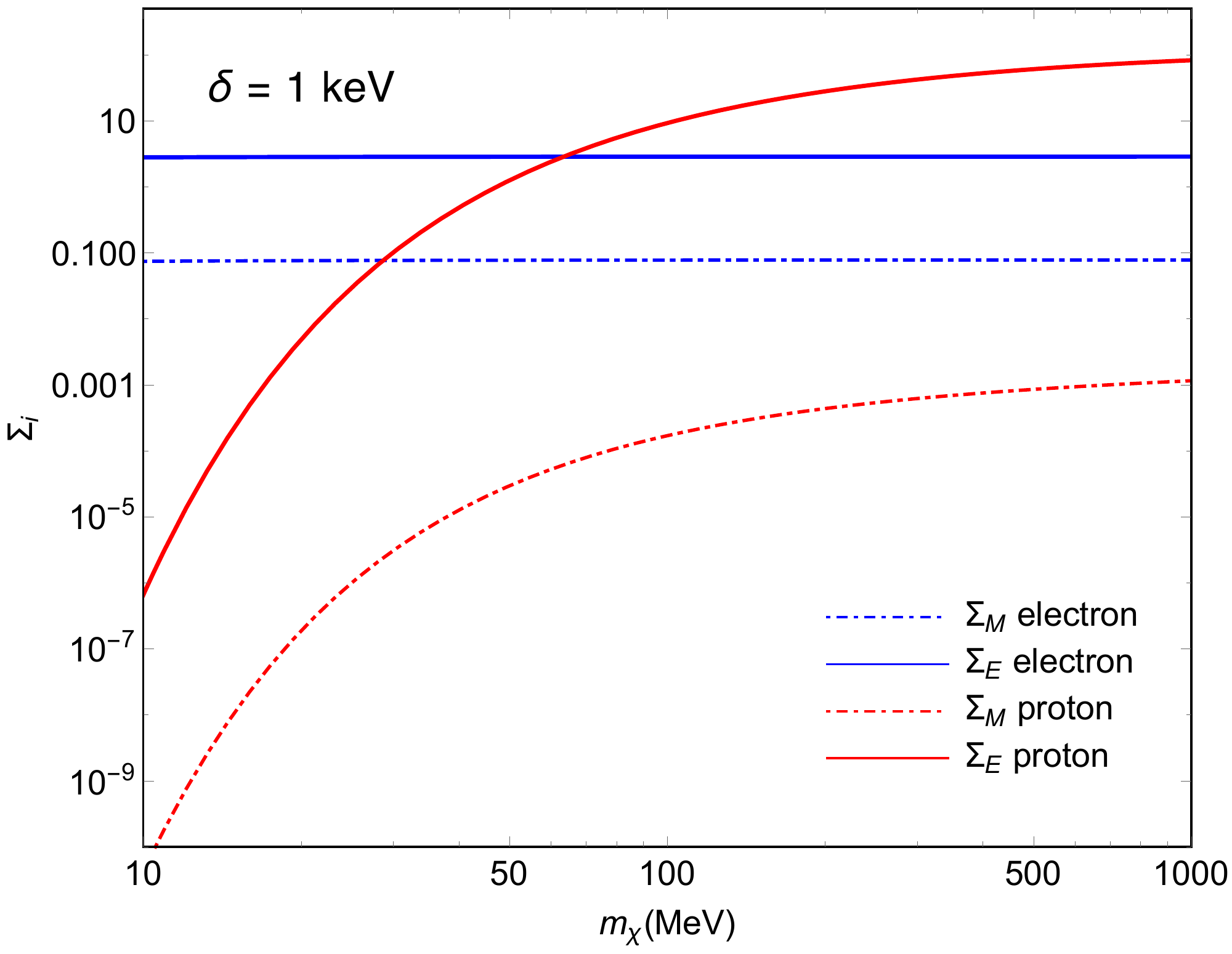}  
\end{center}
\caption{Functions $\Sigma_{M,E}$ defined as in \Eq{eq:SigmaDef} as a function of the DM mass. The parameters are fixed as in \Eq{eq:parameter} and the mass splitting is set to $\delta = 1 \, {\rm keV}$. We show results for magnetic and dipole moments with dot-dashed and solid lines, respectively. Moreover, we separate the contribution from electrons (blue lines) and protons (red lines). } 
\label{fig:Sigma1}
\end{figure}

For the sole purpose of illustration we fix
\be
\begin{split}
v_0 = & \, 10^{-3} c\ , \\  
m_f = & \, (511 \, {\rm keV} , 938 \, {\rm MeV}) \ , \qquad  T_f = 5 \, {\rm keV}  \ .
\label{eq:parameter} 
\end{split}
\ee
The two different values of $m_f$ correspond to electrons and protons, respectively. For this choice of parameters, we plot in Fig.~\ref{fig:Sigma1} the functions $\Sigma_{M,E}$ defined in \Eq{eq:SigmaDef} as a function of the DM mass for $\delta = 1 \, {\rm keV} < T_f$. In particular, we have $\delta < T_f$ and the plasma kinetic energy is able to easily excite $\chi_1$ to the heavier state $\chi_2$. 

As correctly anticipated above, rates from electric dipoles are always larger than those from magnetic dipoles. This effect is due to the small velocity singularity in \Eq{eq:upscatt2}, regularized now by the finite mass splitting, but still capable of enhancing by orders of magnitude the total rate. Another interesting feature of Fig.~\ref{fig:Sigma1} is the drastically different behavior between scattering off of electrons and protons. Scattering off of electrons is independent on $m_\chi$ for the DM mass range under investigation. In contrast, scattering off of protons is much more efficient for larger DM masses. This is a combination of two effects. The Maxwell-Boltzmann suppression for the proton is more severe, as it can be easily seen from the explicit expression for $\mathcal{F}(v_{\rm rel})$ in \Eq{eq:Fvrel}. However, \Eq{eq:Fvrel} is not sensitive to the DM mass, which enters only by setting the threshold velocity. This kinematical limit is given in \Eq{eq:vmin}, where the reduced mass of the DM-fermion system appears in the denominator. Therefore, for $m_\chi$ below the proton mass the threshold velocity is just too high to be thermally accessible.

\subsection{Perseus}

We focus here on the first of two illustrative cases, the Perseus cluster. The rate of dark matter excitation by the interstellar plasma is given by \Eq{eq:PhiDef}. The effective  plasma-times-dark matter number density $\kappa_{\rm eff}$ as defined in \Eq{eq:neffdef} is calculated following \Ref{Carlson:2014lla}, for the observationally relevant angular region. We use a Navarro-Frenk-White density profile \cite{Navarro:1996gj} for the dark matter density distribution in the Perseus cluster of the form
\be
\rho_{\rm DM}(r)=\frac{\rho_0}{\left(\frac{r}{R_s}\right)\left(1+\frac{r}{R_s}\right)^2} \ .
\ee
The parameters are derived using the results of Ref.~\cite{Reiprich:2001zv} with the concentration-mass relation as quoted in Ref.~\cite{Buote:2006kx},
\begin{align}
R_s = & \, 445 \, {\rm kpc} \ , \\
\rho_0 = & \, 0.0217 \, {\rm GeV} / {\rm cm}^3 \ . 
\end{align} 
For the electron density we use the following  double $\beta$-function density profile \cite{Churazov:2003hr}
\be
\begin{split}
n_e(r) = & \, \frac{3.9\times 10^{-2}\ {\rm cm}^{-3}}{\left(1+\left(\frac{r}{80\ {\rm kpc}}\right)^2\right)^{1.8}} + \\ &\frac{4.05\times 10^{-3}\ {\rm cm}^{-3}}{\left(1+\left(\frac{r}{280\ {\rm kpc}}\right)^2\right)^{0.87}} \ .
\end{split}
\ee
The calculation of the integral in \Eq{eq:neffdef} gives, for an angular region between angles of $1^\prime$ and $12^\prime$ of the center of the cluster as used in \Ref{Bulbul:2014sua}, 
\be
\kappa^{\rm Perseus}_{\rm eff} \simeq  1.06\times 10^{18}\ \left(\frac{10 \, {\rm MeV}}{m_{\chi}} \right) \ {\rm cm}^{-5} \ .
\label{eq:neffper}
\ee

\begin{figure}
\begin{center}
\includegraphics[width=0.45\textwidth]{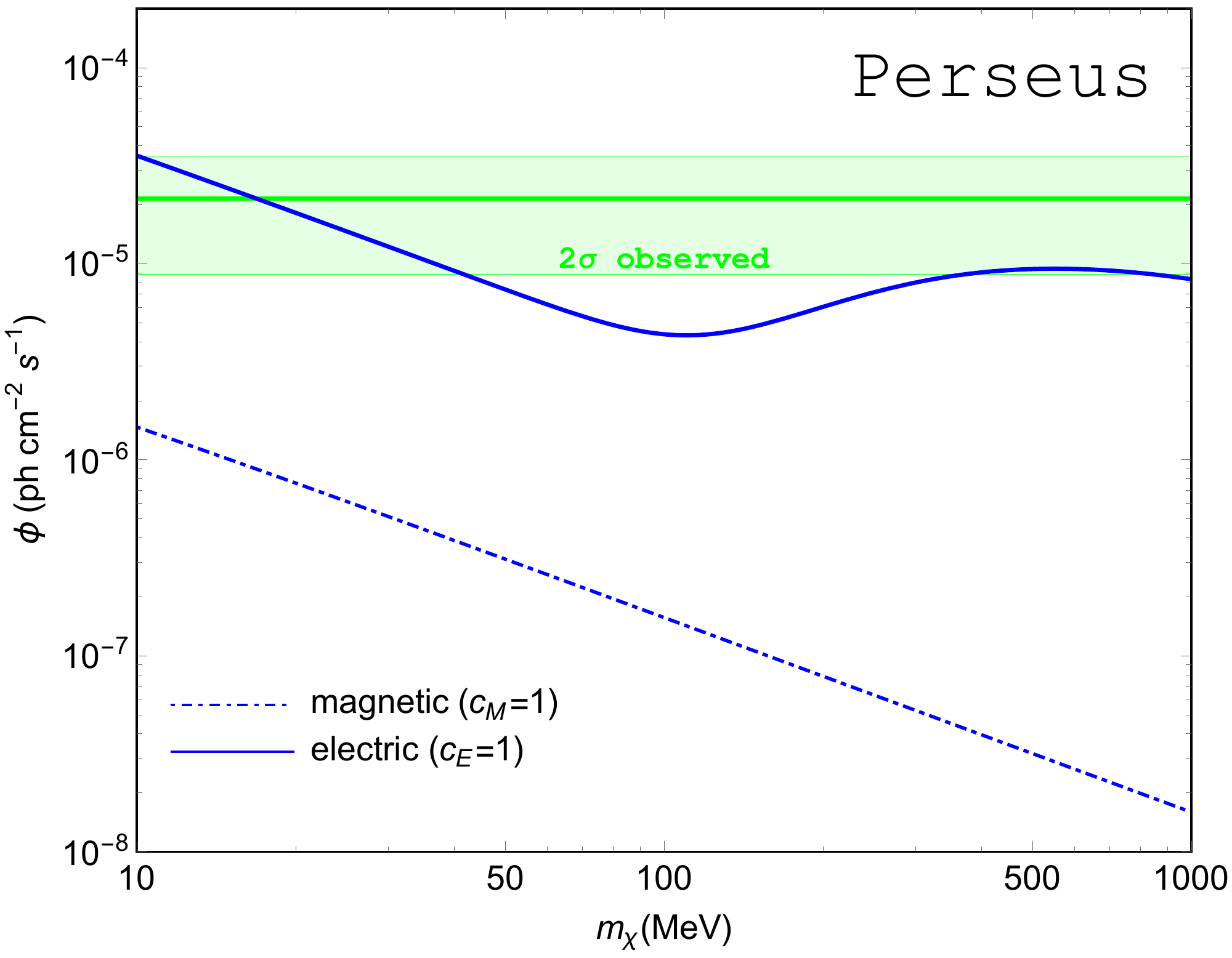}  
\end{center}
\caption{Predicted rate of X-ray photons from the Perseus cluster as a function of the DM mass. We show contributions from both magnetic (dot-dashed lines) and electric (solid lines). Here, we set $\delta = 3.5 \, {\rm keV}$ and $\Lambda = 200 \, {\rm GeV}$. The green band indicates the $2\sigma$ region for the  observed 3.5 keV line flux~\cite{Bulbul:2014sua}.} 
\label{fig:FluxPerseus}
\end{figure}

We fix the virial temperature to~\cite{Reiprich:2001zv}
\be
\left. T_f\right|_{\rm Perseus} = 6.8 \, {\rm keV} \ .
\ee
Moreover, we fix the mass splitting to $\delta = 3.5 \, {\rm keV}$ and the suppression scale to $\Lambda = 200 \, {\rm GeV}$ (the rescaling of the rates with $\Lambda$ is straightforward). 

The results for the fluxes are shown in Fig.~\ref{fig:FluxPerseus}, where we also include the 3.5 keV line flux as quoted in \Ref{Bulbul:2014sua}. The magnetic contribution is negligible. The flux of the X-ray line from electric dipole interactions has an interesting dependence on $m_\chi$: At low DM mass, $m_\chi \simeq 10 \, {\rm MeV}$, the rate is dominated by scattering off of electrons and it falls as $m_\chi^{-1}$ because of the depletion in the DM number density, see \Eq{eq:neffper}. As the DM mass becomes larger than $m_\chi \simeq 100 \, {\rm MeV}$, scattering off of protons becomes efficient for the reasons explained at the end of Section~\ref{sec:ratesgeneral}. This induces a temporary rise in the flux, which eventually starts falling again because of the number density depletion once the proton up-scattering cross section reaches its asymptotic limit. 

\subsection{Galactic Center}

We perform an analogous analysis for the X-ray data from the Galactic center (GC) \Ref{Jeltema:2014qfa}. In this case we conservatively consider a cored Burkert profile \cite{Burkert:1995yz} for the dark matter density (other possibilities are discussed below)
\be
\rho_{\rm DM}(r)=\frac{\rho_0\ R_s^3}{\left(r+R_s\right)\left(r^2+R_s^2\right)} \ .
\ee
We set $\rho_0=2.9$ GeV/cm$^3$ and $R_s=6.0$ kpc \cite{Carlson:2014lla}.
The interstellar gas density is taken to be the sum of the ``thick'' and ``thin'' disk components of \Ref{Cordes:2002wz}, with the parameters provided in Tab.~2 and 3 in \Ref{Cordes:2002wz}, plus an additional ``Galactic Center'' component, for which we use models of the Central Molecular Zone as detailed in Ref.~\cite{Taylor:1993my, Cordes:2002wz, Genzel:2010zy}, and of the hot ionized plasma in the central regions (see e.g. Sec.~3 of \Ref{Genzel:2010zy}).

\begin{figure}
\begin{center}
\includegraphics[width=0.45\textwidth]{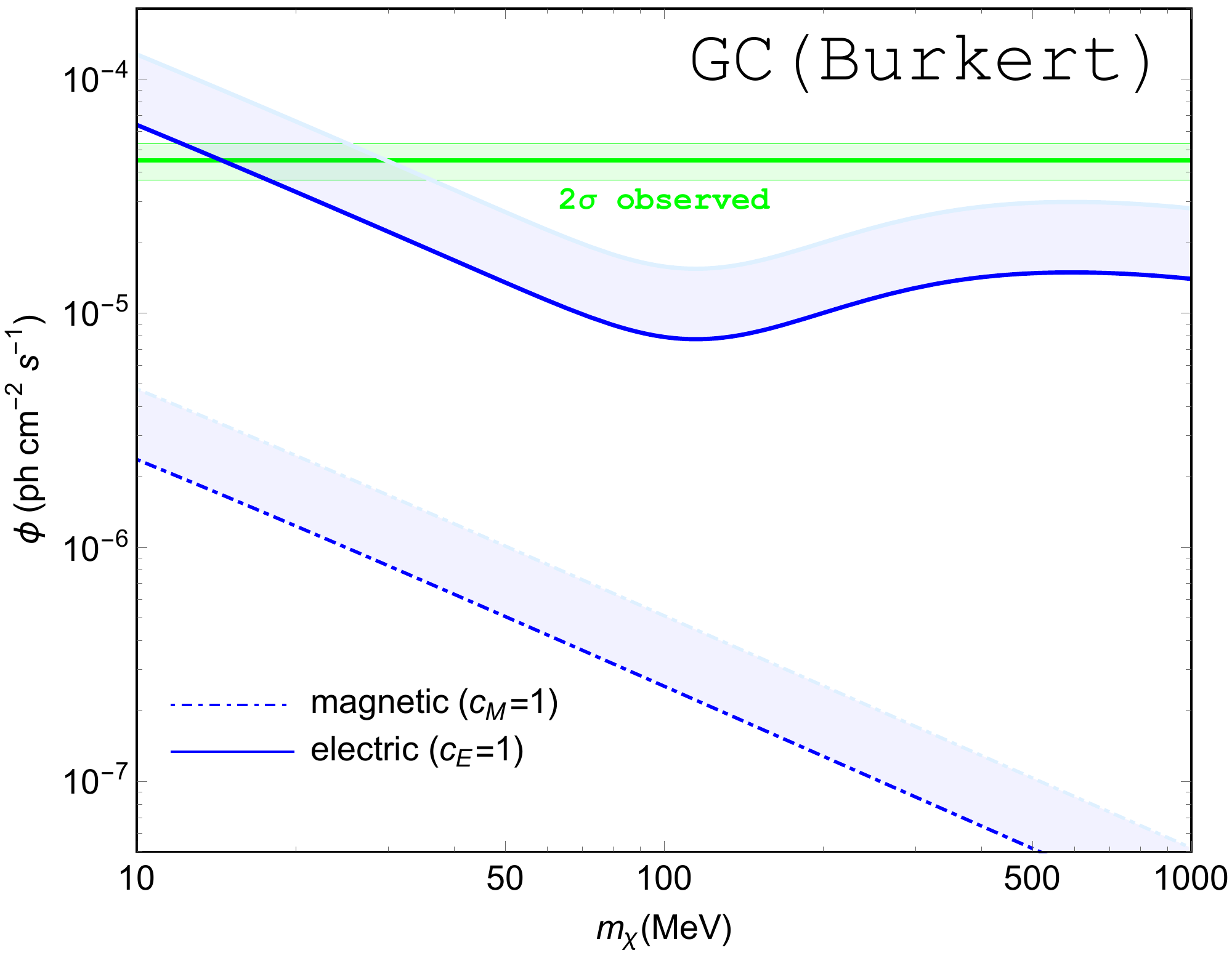}  
\end{center}
\caption{Predicted rate of X-ray photons from the Galactic Center  as a function of the DM mass, with the 2-$\sigma$ range for the detected line flux at energies around 3.5 keV \Ref{Jeltema:2014qfa}. Conventions are as in Fig.~\ref{fig:FluxPerseus}, but we indicate with a blue shading the scatter produced by including different models for the Galactic center gas; in the figure  we again set $\delta = 3.5 \, {\rm keV}$ and $\Lambda = 200 \, {\rm GeV}$.} 
\label{fig:FluxGC2}
\end{figure}

For the GC plasma temperature we consider a multi-temperature model consisting of an admixture of two different temperatures: $T_f = 1 \, {\rm keV}$ with density 4/5 of the total, and the remaining $1/5$ with temperature $T_f = 7 \, {\rm keV}$ \cite{Jeltema:2014qfa}. We checked that employing a single-temperature model does not affect our results significantly. The resulting normalization factor for the excitation rate from the Galactic center is
\be
\kappa^{\rm GC}_{\rm eff}=7.82\times 10^{18}\ \left(\frac{10 \, {\rm MeV}}{m_{\chi}} \right) \ {\rm cm}^{-5},
\ee
with a possible enhancement by a factor of up to 2 depending on the details of the geometry and central density of the Galactic center gas distribution \cite{Taylor:1993my, Cordes:2002wz, Genzel:2010zy}; we illustrate the effect of the Galactic center gas with the light blue shading above the blue lines in fig.~\ref{fig:FluxGC2}.

We present results for the photon flux and comparison with observation (assuming no contribution to the 3.5 keV line from, e.g., K), as detected and reported in \Ref{Jeltema:2014qfa}, in Fig.~\ref{fig:FluxGC2}. 

We also considered other choices for the dark matter density profile. For example, employing a Navarro-Frenk-White density profile \cite{Navarro:1996gj} the predicted rate of X-rays from excitation would be around one order of magnitude larger than with a Burkert profile. This could still be made consistent with observations from Perseus if a fraction of the 3.5 X-ray there were associated with emission from e.g. K XVIII de-excitation.

\section{Discussion and Conclusions}
\label{sec:conclusions}
Fig.~\ref{fig:FluxPerseus} and \ref{fig:FluxGC2} illustrate that our model predicts X-ray fluxes from dark matter thermal plasma excitations from the Galactic center and from Perseus compatible with observations for a range of dark matter masses. For the choice $c_M=c_E=1$ and $\Lambda=200$ GeV, this mass range is in the 10-20 MeV mass range; larger values for $c_E/\Lambda$ would shift the preferred mass range to larger masses; the same would be true if a fraction of the observed X-ray flux at 3.5 keV were to be attributed to other mechanisms than the one under consideration, for example from K ions.

\begin{figure}
\begin{center}
\includegraphics[width=0.45\textwidth]{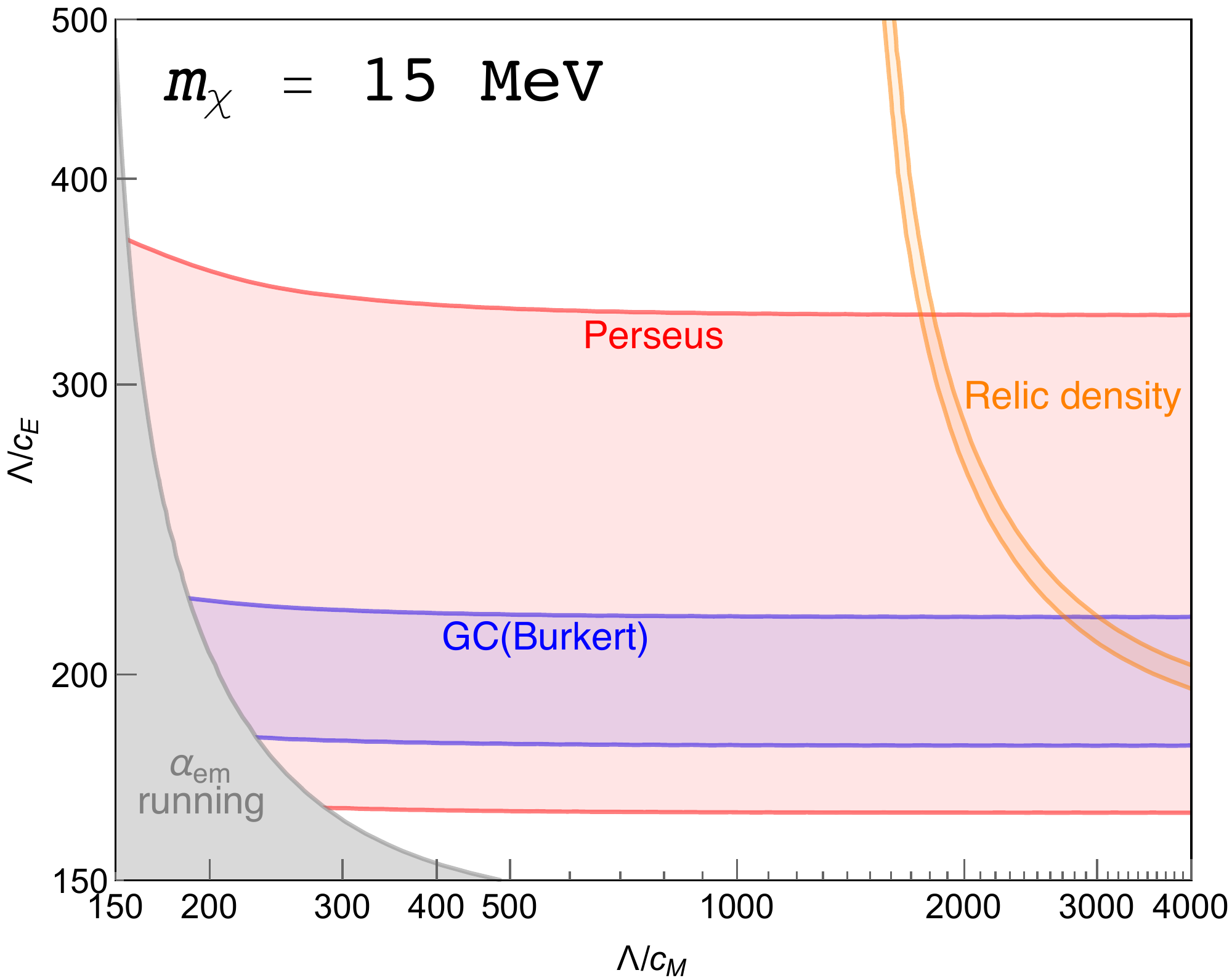}  
\end{center}
\caption{Summary plot for $m_\chi = 15 \, {\rm MeV}$.} 
\label{fig:Summary1}
\end{figure}

Fig.~\ref{fig:Summary1} examines, for a mass $m_\chi=15$ MeV, the parameter space in the $(\Lambda/c_M,\Lambda/c_E)$ plane compatible with X-ray observations {\em and} producing a thermal relic dark matter abundance in accordance with the observed cosmological dark matter density. The red shaded region is compatible with the observed flux from Perseus, and the blue region with the flux detected from the Galactic center (assuming a Burkert density profile, and assuming the whole 3.5 keV line flux is associated only with dark matter de-excitations). Finally, the orange region is compatible with dark matter being entirely produced as a thermal relic, and the grey region is excluded by the running of the electromagnetic coupling.

We thus find that there exists theory parameter space where our models features a thermal relic that explains the 3.5 keV line observations. The preferred ratio of the Wilson coefficients, for the mass choice we made, is $c_E/c_M\sim10-15$, which is not incompatible with generic theoretical expectations for the two parameters being in a similar ballpark.

We emphasize once again that while our model {\em does} provide a consistent explanation to the observational landscape of the 3.5 keV line for thermal relic dark matter, the mass splitting between the two dark sector fermions is a free parameter. As a result, lines could be produced in principle at any energies below the typical temperatures of the plasma (few keV). This opens up a brand new observational window to search for signals from the dark sector. The predicted signals are unique in three respects:
\begin{enumerate}
\item The {\em line-shape} is peculiar: it is a geometric product of the width of the dark matter and of the thermal plasma velocity distribution (see \Eq{eq:Fvrel}); with sufficient energy resolution, such as that expected for the newly launched Hitomi X-ray satellite \cite{2014arXiv1412.1176K}, it will be thus possible to discriminate this mechanism against other sources of X-ray lines such as dark matter decay or plasma de-excitation lines;
\item The {\em morphology} of the predicted emission depends on a unique line-of-sight integral of the product of the plasma density times the dark matter density, in contrast with other models where the signal depends on the dark matter density, density squared, or on magnetic fields;
\item The {\em brightest predicted targets} depend not only on the integral mentioned above, but also on the target's temperature: depending on the dark sector mass splitting, large temperatures might be needed to excite the heavier particle to significant levels; As a result, colder targets might be significantly dimmer than hotter ones.
\end{enumerate}

In principle, absorption of X-ray photons at the $\chi_2-\chi_1$ mass splitting could be present in our model. This possibility has been explored, in a closely related setup to the present one, in Ref.~\cite{Profumo:2006im}. The optical depth corresponding to the choice of parameters we discussed here is, however, extremely small (on the order of $\tau\sim10^{-15}$ using the results of Ref.~\cite{Profumo:2006im}) and absorption features would not be detectable. 

In summary, we have proposed and studied in detail a simple model for the dark sector that offers novel and unique detection prospects. The model consists of two (gauge) ``sterile'' Weyl fermions, odd under a $\ZZ_2$ symmetry, and coupled to the Standard Model via an effective inelastic magnetic and electric dipole operator. The model effectively contains four parameters: the suppression scale for the two dipole operators, and the masses of the two fermions.

We scrutinized in great detail the phenomenology of the model, and concluded that (i) the mass of the lightest state is constrained by cosmology to be heavier than about 10 MeV, and (ii) the strongest constraint on the scale of the effective dipole operators is from corrections to the running of the electromagnetic coupling. We then studied the thermal relic density of the dark matter, and demonstrated that it is driven by the {\em magnetic} dipole term and can easily match the observed universal dark matter density for a broad range of dark matter masses. The model predicts significant production of the heavier fermion  from inelastic collision of the lighter (dark matter) fermion with the thermal plasma in galaxies or clusters of galaxies. This up-scattering rate is driven by the {\em electric} dipole term, and is predicted to source a bright, detectable line through the decay of the heavier fermion to the lighter fermion and a quasi-monochromatic photon.

Remarkably, there exist at least {\em three independent unique observational handles} that could discriminate this scenario from other background processes producing lines or other new physics scenarios: the line's width, its morphology, and the (predictable) dependence on the target's temperature. The energy of the predicted line is essentially a free parameter of the theory, as long as it is at energies comparable or lower than the typical plasma temperature of a given astrophysical target. This model opens a new exciting window for the search of the first non-gravitational manifestation of dark matter as a particle.

\medskip
\section*{Acknowledgments} \noindent  FD is grateful to Howard Haber, Jeremy Mardon and Duccio Pappadopulo for useful discussions. SP is grateful to Eric Carlson for help and consultations on plasma density models. FD, SP and TS are partly supported by the U.S. Department of Energy grant number DE-SC0010107. TS is furthermore supported by a Feodor-Lynen research fellowship sponsored by the Alexander von Humboldt foundation.

\appendix

\section{One-loop Running of the Electromagnetic Coupling}
\label{app:running}

\begin{figure}
\begin{center}
\includegraphics[width=0.3\textwidth]{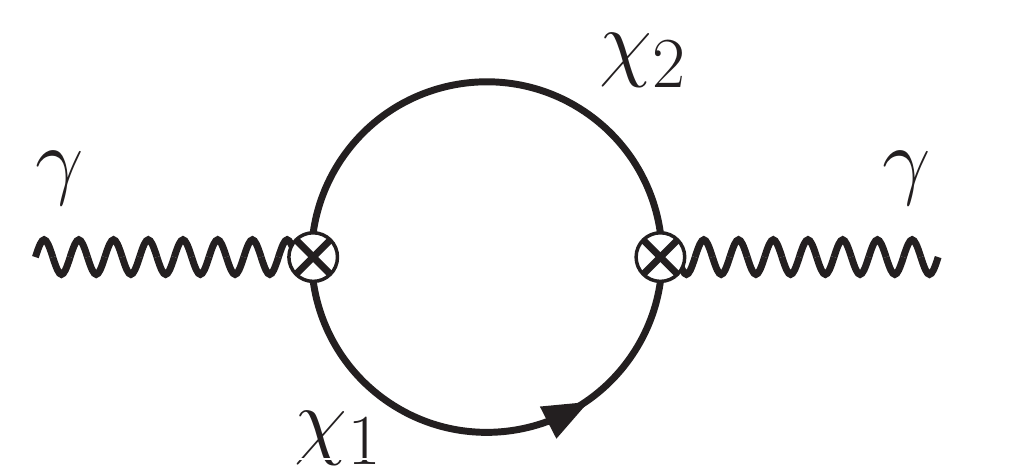}  
\end{center}
\caption{Feynman diagram for the one-loop photon propagator. The crossed circle denotes the insertion of the effective magnetic or electric dipole operator.} 
\label{fig:FeynDiag3}
\end{figure}

In this Appendix we provide the calculation of the one-loop electromagnetic coupling running. We evaluate the contribution to the photon self-energy sketched in Fig.~\ref{fig:FeynDiag3} in the ``Dirac limit" (e.g. by using \Eq{eq:EFT}), justified since we consider $\delta \ll m_\chi$. The photon self-energy results in
\begin{widetext}
\be
i \, \Pi_{\rm dip}^{\mu\nu}(q) = - \frac{1}{\Lambda^2} 
\int \frac{d^d k}{(2\pi)^d} \,  \frac{ {\rm Tr} \left[ \left( \slashed{k} + m_\chi \right) q_\alpha \Sigma^{\alpha\mu} (c_M + i c_E \gamma^5) \left( \slashed{k} + \slashed{q} + m_\chi \right) q_\beta \Sigma^{\beta\nu} (c_M + i c_E \gamma^5) \right]}{(k^2 - m_\chi^2) ((k+q)^2 - m_\chi^2)}
 \ ,
\ee
\end{widetext}
where we regularize the UV behavior by computing the loop integral in $d$ space-time dimensions. The calculation proceeds by introducing the Feynman parameters and performing the change of integration variable $l = k + q x$
\begin{widetext}
\be
i \, \Pi_{\rm dip}^{\mu\nu}(q) = - \frac{1}{\Lambda^2}  \int_0^1 dx
\int \frac{d^d l}{(2\pi)^d}  \frac{ {\rm Tr} \left[\left( \slashed{l} - \slashed{q} x + m_\chi \right) q_\alpha \Sigma^{\alpha\mu} (c_M + i c_E \gamma^5) \left( \slashed{l} + \slashed{q}(1-x) + m_\chi \right) q_\beta \Sigma^{\beta\nu} (c_M + i c_E \gamma^5) \right]}{(l^2 - \Delta)^2} \ ,
\label{eq:loop}
\ee
\end{widetext}
where we also introduced
\be
\label{eq:Deltadef} \Delta =  m_\chi^2 - q^2 x (1 - x) \ .
\ee
All the terms with an odd power of $l$ in the numerator of \Eq{eq:loop} vanish by parity consideration. The Dirac trace of the surviving terms is straightforward. Consistently with electromagnetic gauge invariance, the expression we find is transverse, $\Pi_{\rm dip}^{\mu\nu}(q) = \left(q^2 g^{\mu\nu} - q^\mu q^\nu \right) \Pi_{\rm dip}(q^2)$, where the quantity $\Pi(q)$ results in
\begin{widetext}
\be
\Pi_{\rm dip}(q^2) =  i \frac{4}{\Lambda^2}  \int_0^1 dx
\int \frac{d^d l}{(2\pi)^d}  \left[ \frac{(c_M^2 + c_E^2) q^2 x (1 - x) + (c_M^2 - c_E^2) m_\chi^2}{(l^2 - \Delta)^2} +   \frac{ \left( 1 - \frac{4}{d} \right) (c_M^2 + c_E^2) \, l^2}{(l^2 - \Delta)^2}  \right]\ .
\label{eq:loop2}
\ee
\end{widetext}
The term proportional to $l^2$ vanishes in $d=4$, and it can be explicitly checked that it has no $d = 4$ pole in dimensional regularization. However, it has a finite piece that we keep in our analysis.

The next step is to compute the loop integral in \Eq{eq:loop2} in $d = 4 - 2 \epsilon$ space-time dimensions. The two terms can be evaluated by using the general results
\begin{align}
\label{eq:loopint1} \int \frac{d^d l}{(2\pi)^d} \frac{1}{{(l^2 - \Delta)^2} } = & \, i \,  \mathfrak{L}_0 \ , \\
\label{eq:loopint2} \int \frac{d^d l}{(2\pi)^d} \frac{l^2}{{(l^2 - \Delta)^2} } = & \, i \, \frac{\Delta}{1 - 2/d} \; \mathfrak{L}_0 \ .
\end{align}
Here, we define 
\be
\mathfrak{L}_0 \equiv \frac{\Gamma(2 - d/2)}{(4 \pi)^{d/2} \, \Delta^{2 - d/2}} \ ,
\label{eq:L0def}
\ee
with $\Gamma(x)$ the Euler function. Upon Taylor-expanding around 4 dimensions, with expansion parameter $ \epsilon = (4 - d) / 2$, we find
\be
\Gamma(2 - d/2) = \Gamma(\epsilon) = \frac{1}{\epsilon} - \gamma_E + \mathcal{O}(\epsilon)  \ ,
\ee
where $\gamma_E \simeq 0.577$ is the Euler-Mascheroni constant. Therefore in $d = 4 - 2 \epsilon$ dimensions, up to $\mathcal{O}(\epsilon) $ terms, the quantity $\mathfrak{L}_0$ defined in \Eq{eq:L0def} reads
\be
\mathfrak{L}_0 = \frac{1}{16 \pi^2} \left[ \frac{1}{\epsilon} + \ln\left(\frac{4 \pi \, e^{-\gamma_E}}{m_\chi^2 - q^2 x (1 - x)}\right) \right]  \ .
\ee
Here, we have restored the explicit functional form for $\Delta$ as given in \Eq{eq:Deltadef}.

After computing the loop integral as we have just described, the photon self-energy in \Eq{eq:loop2} becomes
\begin{widetext}
\be
\begin{split}
\Pi_{\rm dip}(q^2) = & \,  - \frac{1}{\epsilon} \, \frac{1}{4 \pi^2 \Lambda^2}  \left[ (c_M^2 + c_E^2) \frac{q^2}{6} + (c_M^2 - c_E^2) m_\chi^2  \right] + \frac{ (c_M^2 + c_E^2)}{4 \pi^2 \Lambda^2}  \left(m_\chi^2 - \frac{1}{6} q^2 \right) \\ &
- \frac{1}{4 \pi^2 \Lambda^2}  \int_0^1 dx \left[ (c_M^2 + c_E^2) q^2 x (1 - x) + (c_M^2 - c_E^2) m_\chi^2  \right] \ln\left(\frac{4 \pi \, e^{-\gamma_E}}{m_\chi^2 - q^2 x (1 - x)}\right)  \ ,
\end{split}
\label{eq:loop3}
\ee
\end{widetext}
where we only keep terms surviving after the $\epsilon \rightarrow 0$ limit.

The one-loop photon self-energy in \Eq{eq:loop3} has a $1 / \epsilon$ divergence that needs to be renormalized. The pole proportional to the squared DM mass is taken care of by a renormalization of the photon kinetic term, as it happens in regular QED. However, the pole proportional to $q^2$ requires the introduction of a new dimension 6 operator
\be
\mathcal{O}_6 = \frac{1}{4} F^{\mu \nu} \Box F_{\mu\nu} \ .
\label{eq:O6}
\ee
The renormalized photon self-energy is given by the one-loop contribution in \Eq{eq:loop3} summed to the two infinite counterterms mentioned above
\be
\Pi_{\rm dip}^{\rm ren}(q^2)  = \Pi_{\rm dip}(q^2) - \delta_4 - \delta_6 \, q^2 \ ,
\label{eq:PiFull}
\ee
where $\delta_4$ and $\delta_6$ are the counterterms for the photon kinetic term and the dimension 6 operator in \Eq{eq:O6}, respectively.

The next step is to go from the full self-energy in \Eq{eq:PiFull} to the running electromagnetic coupling. We compare our prediction with Bhabha-scattering data. As is well known~\cite{GellMann:1954fq}, the running coupling is given by
\be
\alpha_{\rm em}(q^2) = \frac{\alpha_0}{1 - \Pi(q^2)} \ ,
\label{eq:alphaemDEF}
\ee
where $\alpha_0$ is the fine structure constant and $\Pi(q^2)$ is the full photon self-energy, which is given by the sum of the SM contribution and the renormalized dipole contribution in \Eq{eq:PiFull}. The SM contribution to the QED vacuum polarization is well known~\cite{Steinhauser:1998rq,Cabibbo:1961sz,Burkhardt:2001xp,Burkhardt:2011ur}. All we have left to do is to add the DM contribution, but before doing so we have to fix the counterterms $\delta_4$ and $\delta_6$ appearing in \Eq{eq:PiFull}. We start from fixing $\delta_4$, and we do that by requiring that $\alpha_{\rm em}(q^2 = 0) =  \alpha_0$, or in other words we choose $\delta_4$ to make sure that $\Pi_{\rm dip}^{\rm ren}(q^2 = 0) = 0$ as follows
\begin{widetext}
\be
\begin{split}
\Pi_{\rm dip}^{\rm ren}(q^2) = & \,  - q^2 \frac{c_M^2 + c_E^2}{24 \pi^2 \Lambda^2}  \left( \frac{1}{\epsilon} + 1\right) - \frac{(c_M^2 - c_E^2) m_\chi^2}{4 \pi^2 \Lambda^2}  \int_0^1 dx  \ln\left(\frac{m_\chi^2}{m_\chi^2 - q^2 x (1 - x)}\right) \\ &
- q^2 \frac{c_M^2 + c_E^2}{4 \pi^2 \Lambda^2}  \int_0^1 dx \; x (1 - x) \ln\left(\frac{4 \pi \, e^{-\gamma_E}}{m_\chi^2 - q^2 x (1 - x)}\right) - \delta_6 \, q^2 \ .
\end{split}
\ee
\end{widetext}
The renormalization condition at $q^2 = 0$ cannot fix the counterterm $\delta_6$, because its effects vanish in the static limit. Thus in order to fix $\delta_6$ we need another experimental input. At energy scales much smaller than the DM mass we consider, namely $ - q^2 \ll (10 \, {\rm MeV})^2$, the DM particles in the loop cannot affect the gauge coupling running. We evaluate \Eq{eq:alphaemDEF} in such regime, and find the $\delta_6$ we need in order to have only the SM contribution. The final renormalized DM contribution to the photon self-energy reads 
\begin{widetext}
\be
\Pi_{\rm dip}^{\rm ren}(q^2) = - m_\chi^2 \frac{(c_M^2 - c_E^2)}{4 \pi^2 \Lambda^2}  \int_0^1 dx  \ln\left(\frac{m_\chi^2}{m_\chi^2 - q^2 x (1 - x)}\right)
- q^2 \frac{c_M^2 + c_E^2}{4 \pi^2 \Lambda^2}  \int_0^1 dx \; x (1 - x) \ln\left(\frac{m_\chi^2}{m_\chi^2 - q^2 x (1 - x)}\right) \ .
\label{eq:PiRenDip2}
\ee
\end{widetext}

\section{A Simple Formula for the Thermal Average of the Excitation Rate}
\label{app:sigmav}

In this Appendix we derive the simple expression for the thermal average given in \Eq{eq:sigmavrel}. We start from the definition in \Eq{eq:sigmavdef}, which involves six integrals. It is convenient to change integration variables
\begin{align}
\bs V = & \, \frac{\bs v_\chi + \bs v_f}{2} \ , \\
\bs v_{\rm rel} = & \, \bs v_\chi - \bs v_f \ .
\end{align}
The up-scattering cross section is a Lorentz-invariant quantity that depends only on the CM energy $\sqrt{s}$ of the collision. In the non-relativistic limit we have
\be
s = \left( m_\chi + m_f + \frac{1}{2} \frac{m_\chi m_f}{m_\chi + m_f} v_{\rm rel}^2 \right)^2 \ ,
\ee
therefore the cross section can only depend on the absolute value of the relative velocity between the two particles, as correctly reproduced in \Eq{eq:sigmavrel}.

The inelastic collision initial state is characterized by three quantities: the magnitude of the two velocities and the angle between them. The integration in the new variables can be performed in polar coordinates, i.e.
\be
d^3 V d^3 v_{\rm rel} \;\; \rightarrow \;\;  4 \pi \times 2 \pi \times   dV \,  d v_{\rm rel} \, d \cos\gamma \, V^2 v^2_{\rm rel} \ ,
\ee
where we define $\gamma$ as the angle between $\bs V$ and $\bs v_{\rm rel}$. Upon expressing the thermal average as in \Eq{eq:sigmavrel}, the function $\mathcal{F}(v_{\rm rel})$ reads
\be
\begin{split}
\mathcal{F}(v_{\rm rel}) \equiv & \, 8 \pi^2 v_{\rm rel}^3 \int_0^\infty  dV \, V^2 \int_{-1}^1 d \cos\gamma \; \times \\ &
f_{\rm DM}(V, v_{\rm rel}, \cos\gamma) f_f(V, v_{\rm rel}, \cos\gamma) \ .
\end{split}
\ee
The distribution functions can be easily expressed as a function of $V$ and $v_{\rm rel}$ by using the relations
\begin{align}
v_\chi^2 = \bs v_\chi \cdot \bs v_\chi = & \, V^2 + \frac{v_{\rm rel}^2}{4} + V v_{\rm rel} \cos\gamma \ , \\
v_f^2 = \bs v_f \cdot \bs v_f = & \,  V^2 + \frac{v_{\rm rel}^2}{4} - V v_{\rm rel} \cos\gamma  \ .
\end{align}

All we have to do to get to \Eq{eq:Fvrel} is to introduce the explicit form of the distribution functions. We integrate over the angle $\gamma$ by using the relation
\be
\begin{split}
&  \int_{-1}^1 d \cos\gamma \exp\left[ - \left(a_\chi - a_f \right) V v_{\rm rel} \cos\gamma \right] =  \\ &
 2 \, \frac{\sinh\left[\left(a_\chi - a_f \right) V v_{\rm rel} \right]}{\left(a_\chi - a_f \right) V v_{\rm rel}} \ ,
\end{split}
\ee
and consequently the function $\mathcal{F}(v_{\rm rel})$ results in
\be
\begin{split}
\mathcal{F}(v_{\rm rel}) = & \,  \frac{16 v_{\rm rel}^2 \, a_\chi^{3/2} a_f^{3/2} \, \exp\left[ - \left(a_\chi + a_f \right) \frac{v_{\rm rel}^2}{4}  \right] }{\pi \left(a_\chi - a_f \right)} \times  \\ & 
\int_0^\infty  dV \, V  \exp\left[ - \left(a_\chi + a_f \right) V^2 \right]  \times \\ &
\sinh\left[\left(a_\chi - a_f \right) V v_{\rm rel} \right]\ .
\end{split}
\ee
Finally, the integration over $V$ can be performed by using 
\be
\begin{split}
& \int_0^\infty  dV \, V \,  \exp\left[ - \alpha V^2 \right] \; \sinh\left[\beta V \right] = \\ &
\frac{\pi^{1/2}}{4} \frac{\beta}{\alpha^{3/2}} \exp\left[ \frac{\beta^2}{4 \alpha}\right]  \ ,
\end{split}
\ee
with the final expression for $\mathcal{F}(v_{\rm rel})$ as given in \Eq{eq:Fvrel}.

\bibliography{ExcitingDipole}

\end{document}